\pdfoutput=1

\newif\ifarxiv
\arxivtrue

\documentclass[sigconf]{acmart}

\copyrightyear{2024}
\acmYear{2024}
\setcopyright{acmlicensed}
\acmConference[CCS '24]{Proceedings of the 2024 ACM SIGSAC Conference on Computer and Communications Security}{October 14--18, 2024}{Salt Lake City, UT, USA}
\acmBooktitle{Proceedings of the 2024 ACM SIGSAC Conference on Computer and Communications Security (CCS '24), October 14--18, 2024, Salt Lake City, UT, USA}
\acmDOI{10.1145/3658644.3690194}
\acmISBN{979-8-4007-0636-3/24/10}

\usepackage[utf8]{inputenc}
\usepackage[T1]{fontenc} %
\usepackage{textcomp} %

\graphicspath{{graphics/}}
\usepackage{amsfonts}
\usepackage{amsthm} %
\usepackage{mathtools} %
\usepackage{algorithmic} %

\usepackage{subcaption}

\usepackage[capitalise,noabbrev]{./cleveref} %
\usepackage{url}
\usepackage{enumitem}

\newcommand{\dpeps}{\epsilon}
\newcommand{\dpdelta}{\delta}
\newcommand{\sai}{Split-AI}

\newcommand{\saiModels}{K}
\newcommand{\cifar}{CIFAR-10}

\newcommand{\rlthresh}{{\ell_\tau}}

\newcommand{\tpr}{\text{TPR}}
\newcommand{\fpr}{\text{FPR}}
\newcommand{\varfpr}{\alpha}

\DeclareMathOperator{\mistat}{\mathcal{S}}

\newcommand{\numShadow}{S}
\newcommand{\numCanaries}{C}
\DeclarePairedDelimiter\abs{\lvert}{\rvert}
\newcommand{\mithresh}{{T}}
\newcommand{\varmithresh}{t}

\makeatletter
\newcommand{\myparagraph}{%
  \@startsection{paragraph}{4}%
  {\z@}{1ex \@plus 1ex \@minus .2ex}{-1em}%
  {\noindent\normalfont\normalsize\bfseries\maybe@addperiod}%
}
\newcommand{\maybe@addperiod}[1]{%
  #1\@addpunct{.}%
}
\newcommand{\softparagraph}[1]{%
{\noindent\emph{#1}}
}
\makeatother

\DeclareMathOperator*{\argmax}{arg\,max}

\newif\ifshownotes
\shownotesfalse
\usepackage{xcolor}

\newcommand{\crefApp}[1]{%
\ifarxiv%
\cref{#1}
\else%
\cref{app:see_more}
\fi%
}

\newcommand{\captitle}[1]{\emph{#1}}
\newcommand{\subcaptitle}[1]{\emph{#1}}

\begin{document}

\title{Evaluations of Machine Learning Privacy Defenses are Misleading}

\author{Michael Aerni}
\orcid{0000-0003-3276-2678}
\authornote{Both authors contributed equally to this research.}
\affiliation{
    \institution{ETH Zurich}
    \country{Switzerland}
}

\author{Jie Zhang}
\orcid{0009-0008-4670-9038}
\authornotemark[1]
\affiliation{
    \institution{ETH Zurich}
    \country{Switzerland}
}

\author{Florian Tramèr}
\orcid{0000-0001-8703-8762}
\affiliation{
    \institution{ETH Zurich}
    \country{Switzerland}
}

\begin{abstract}
    \emph{Empirical defenses} for machine learning privacy
    forgo the provable guarantees of differential privacy
    in the hope of achieving higher utility
    while resisting realistic adversaries.
    We identify severe pitfalls in existing empirical privacy evaluations
    (based on membership inference attacks)
    that result in misleading conclusions.
    In particular, we show that prior evaluations
    fail to characterize the privacy leakage of the \emph{most vulnerable samples},
    use \emph{weak attacks},
    and avoid comparisons with \emph{practical differential privacy baselines}.
    In 5 case studies of empirical privacy defenses,
    we find that prior evaluations underestimate privacy leakage by an order of magnitude.
    Under our stronger evaluation,
    none of the empirical defenses we study are competitive
    with a properly tuned, high-utility DP-SGD baseline
    (with vacuous provable guarantees).
\end{abstract}



\maketitle

\newlength{\figonecolwidth}
\setlength{\figonecolwidth}{230pt}
\newlength{\figtwocolwidth}
\setlength{\figtwocolwidth}{480pt}
\newlength{\figreducedwidth}
\setlength{\figreducedwidth}{166pt}

\newcommand{\canaryChoicesTable}{%
\begin{table}[h]
    \renewcommand{\arraystretch}{1.25}
    \caption{\captitle{Canaries have to be adapted to the defense.}
    Mislabeled samples are strong canaries for defenses that rely on a standard supervised learning paradigm, but fail to mimic leakage of the most vulnerable samples for SELENA or SSL.}
    \label{tab:canaries}
    \begin{tabular}{@{}lp{3cm}@{}}
    \toprule
    Method & Canary Choice \\ \midrule
    HAMP & mislabeled samples \\
    RelaxLoss & mislabeled samples \\
    SELENA & mislabeled duplicates \\
    SSL (SimCLR and MoCo)& OOD data (ImageNet) \\
    DFKD & mislabeled samples \\
    DP-SGD & mislabeled samples, or OOD data (ImageNet), or uniform data \\ \bottomrule
    \end{tabular}
\end{table}
}

\section{Introduction}
Machine learning models can memorize sensitive information from their training data,
enabling privacy attacks such as membership inference~\citep{shokri2016membership}
and data extraction~\citep{carlini2020extracting}.
Training with differential privacy~\cite{dwork2006calibrating}---in particular
with DP-SGD~\citep{abadi2016deep}---provides provable protection against such attacks.
Yet, achieving strong guarantees with good utility
remains a challenge~\citep{feldman2020memorization}.
This has led to growing interest in \emph{empirical} privacy defenses,
which might offer a better privacy-utility tradeoff against practical attacks,
but no formal guarantees~\citep{nasr2018machine,jia2019memguard,yang2020defending,tang2022selena,salem2018mlleaks,chen2022relaxloss,chen2024hamp}.

\begin{figure}[t]
    \centering
    \includegraphics[width=\figonecolwidth]{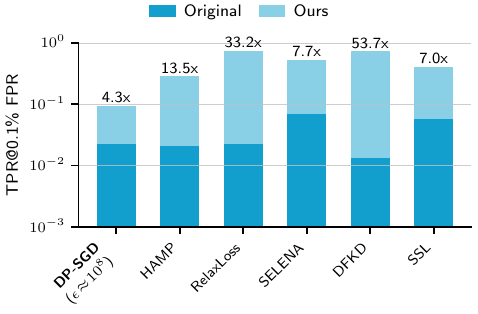}
    \vspace{-0.5em}
    \caption{
    \captitle{Empirical privacy evaluations provide a false sense of security.}
    We study five heuristic defenses and a properly tuned DP-SGD baseline
    that all achieve $\geq$ 88\% accuracy on \cifar{}.
    We first run a standard membership inference evaluation
    and report the attack's TPR at a low FPR across the dataset (following~\citep{carlini2022membership}).
    Our new evaluation methodology,
    which adapts the attack to each defense and targets the least-private samples,
    reveals an order-of-magnitude higher privacy leakage.
    Our DP-SGD baseline provides better privacy (at similar utility)
    than all the empirical defenses.
    \vspace{-1em}
    }
    \label{fig:teaser}
\end{figure}

Most evaluations of such empirical defenses for private machine learning
use \emph{membership inference attacks}~\citep{shokri2016membership}
as the canonical approach to obtain a bound on privacy leakage.
Under the notion of membership privacy,
many heuristic defenses claim to achieve a better privacy-utility tradeoff than DP-SGD
against state-of-the-art attacks~\citep{jia2019memguard,tang2022selena,chen2022relaxloss,chen2024hamp}.
However, we find that such empirical evaluations can be severely misleading:

\begin{enumerate}[leftmargin=8mm, itemsep=3pt]
    \item Current membership inference evaluations fail to reflect a model's privacy
    \emph{on the most vulnerable data},
    and instead aggregate the attack success over a population.
    \textbf{But privacy is not an average-case metric!~\citep{DPorg-average-case-dp}}
    We show that \emph{a blatantly non-private defense
    that fully leaks one training sample passes existing evaluations}
    (even with recent proposals to report an attack's true positive rate
    at low false positive rates~\citep{carlini2022membership}).

    \item Many defenses apply either a weak inference attack
    that does not reflect the current state-of-the-art~\citep{carlini2022membership,ye2022enhanced},
    or fail to properly \emph{adapt} the attack to account
    for unusual defense components or learning paradigms.
    This issue is reminiscent of well-known pitfalls
    for non-adaptive evaluations
    of machine learning robustness~\citep{athalye2018obfuscated,tramer2020adaptive}.

    \item Empirical defenses are typically compared to
    \emph{weak DP-SGD baselines}~\citep{jia2019memguard,tang2022selena,chen2022relaxloss,chen2024hamp}
    with utility below the state-of-the-art.
\end{enumerate}

To address the first issue, we introduce an efficient evaluation methodology
that accurately reflects a defense's privacy on the most vulnerable data points.
Inspired by work on worst-case privacy auditing~\citep{carlini2019secret,jagielski2020auditing},
we inject canary samples that mimic the most vulnerable data,
and focus our audit on those canaries only.

Then, for five representative empirical defenses,
we design adaptive membership inference attacks
based on LiRA~\citep{carlini2022membership}, the state-of-the-art,
and evaluate privacy using our new methodology.
As \cref{fig:teaser} shows, we reveal much stronger privacy leakage
and a completely different ranking
than the original evaluations suggest.
None of the five defenses provide effective protection against properly adapted attacks
targeted at the most vulnerable samples.

Finally, we show that none of these defenses are competitive
with a strong DP-SGD baseline.
By using state-of-the-art improvements to the
original DP-SGD algorithm (e.g.,~\cite{de2022unlocking}),
and by tuning hyperparameters to achieve both high utility and high \emph{empirical} privacy
(at the expense of meaningful provable guarantees),
we obtain a \emph{better empirical privacy-utility tradeoff} than all other defenses.

Our work adds to the growing literature on pitfalls
in evaluations of ML privacy defenses~\citep{choquette2021label,tramer2022debugging,carlini2022no,kaplan2024cautionary,xiao2024dataenhancement}.
We aim to provide a more principled evaluation framework,
and an overview of pitfalls and misconceptions in existing evaluations.
To promote reproducible research,
we release all code for our evaluation methodology
and our implementation of each empirical defense we study.\footnote{
\url{https://github.com/ethz-spylab/misleading-privacy-evals}
}

\section{Preliminaries and Related Work}

\subsection{Privacy Attacks}
Machine learning models can memorize parts of their training data,
enabling various privacy attacks.
\emph{Membership inference}---which we focus on in this work---corresponds to
the most general form of data leakage:
inferring whether a particular data point
was part of a model's training set~\citep{shokri2016membership}.
Stronger attacks such as \emph{attribute inference}~\citep{fredrikson2015model}
or \emph{data extraction}~\citep{carlini2019secret,carlini2020extracting}
aim to recover partial or full training samples by interacting with a model.

\myparagraph*{Membership inference attacks}
In a membership inference attack,
an adversary tries to guess whether some target sample
was in the training data of a machine learning model.

Most membership inference attacks follow a common blueprint:
For a trained model $f$ and target sample $x$,
the attack computes a \emph{score} $\mistat(f; x)$,
typically related to the training loss function
(e.g., the sample's negative cross-entropy loss).
Then, the attack guesses that $x$ is a member if $\mistat(f; x) > \varmithresh$ for some threshold $\varmithresh$.

Early membership inference attacks use a \emph{global threshold} $\tau$ for all samples~\citep{yeom2018privacy, shokri2016membership}.
A number of follow-up works highlight that a global threshold is suboptimal, as some samples are harder to learn than others~\citep{sablayrolles2019white, carlini2022membership,watson2021importance, ye2022enhanced}. Thus, calibrating the attack threshold to each sample
greatly improves membership inference.

\myparagraph*{The Likelihood Ratio Attack (LiRA)}
In this work, we build upon the LiRA framework of \citet{carlini2022membership},
which frames membership inference as a hypothesis testing problem.
Given a target sample $x$, LiRA models the score distributions
under the hypotheses that $x$ is a member of the training data,
and that $x$ is a non-member.
Given the score of the victim model $f$ on $x$,
the attack then applies a \emph{likelihood ratio test}
to distinguish between the two hypotheses.

To estimate the score distributions, LiRA trains multiple \emph{shadow models}~\citep{shokri2016membership}
by repeatedly sampling a training set $D$
(from the same or similar distribution as the training set of $f$),
and training models $f_{\textrm{out}}$ on $D$
and $f_{\textrm{in}}$ on $D\cup\{x\}$.
Given sufficiently many shadow models,
LiRA fits two Gaussians
$\mathcal{N}(\mu_{x,\textrm{in}}, \sigma_{x,\textrm{in}}^2)$ and
$\mathcal{N}(\mu_{x,\textrm{out}}, \sigma_{x,\textrm{out}}^2)$
to the scores from the ``in'' and ``out'' models on the target sample $x$.
Finally, LiRA applies a standard Neyman–Pearson test
to determine whether the observed score $\mistat(f; x)$
from the victim model $f$ is more likely if $x$ is a member or a non-member:
\[
\mathcal{A}(f, x) \coloneqq
\frac{
    \mathcal{N}(\texttt{score}_x \mid \mu_{x,\textrm{in}}, \sigma_{x,\textrm{in}}^2)
}{
    \mathcal{N}(\texttt{score}_x \mid \mu_{x,\textrm{out}}, \sigma_{x,\textrm{out}}^2)
}, \enspace \text{where } \texttt{score}_x = \mistat(f; x) \,.
\]

As an optimization,
if the training algorithm relies on data augmentation,
we can query the model on multiple augmentations of the target input $x$.
LiRA then fits a multivariate Gaussian distribution to the corresponding scores.
Follow-up work also considers querying models on additional samples
(e.g.,~\cite{wen2022coalmine}),
and improving the attack's computational efficiency
(e.g.,~\cite{zarifzadeh2024lchpmembership}).

\subsection{Privacy Defenses}
Defenses against privacy attacks,
in particular against membership inference, fall into two broad categories.

\myparagraph*{Provable defenses}
A differential privacy (DP)~\cite{dwork2006calibrating} machine learning algorithm
provably bounds the success of typical privacy attacks.
Differentially private models are often trained
using the DP-SGD algorithm~\cite{abadi2016deep},
which protects each individual training step
by clipping and noising per-sample gradients.
For many tasks,
achieving strong provable privacy (e.g.,~$\dpeps \approx 1$) with DP-SGD
requires a large noise magnitude,
which deteriorates model utility.

If some \emph{public} data is available,
better privacy-utility tradeoffs are possible
with techniques such as PATE~\cite{papernot2016semi},
or public pretraining followed by private fine-tuning~\cite{tramer2020differentially,de2022unlocking,pinto2024PILLAR}.
This paper focuses on the strict privacy setting,
where all training data has to be protected.

\myparagraph*{Empirical defenses}
Due to the high utility cost of provable privacy guarantees,
many heuristic defenses
aim for \emph{empirical} privacy against realistic attacks.
Existing heuristic defenses rely on techniques
such as adversarial training~\citep{nasr2018machine},
modifications to a model's loss or confidence~\citep{jia2019memguard,chen2022relaxloss,chen2024hamp, yang2020defending},
or indirect access to private features or labels
(e.g., through distillation~\citep{tang2022selena},
self-supervised learning~\citep{moco, simclr},
or synthetic data generation~\citep{lopes2017data,yin2020dreaming,fang2022up, dong2022privacy}).

\subsection{Empirical Privacy Evaluation}
\label{ssec:prelims_evals}

\myparagraph*{Membership inference evaluations}
Membership inference (MI) attacks and defenses
are typically evaluated on a dataset containing the victim model's training data and an equal number of non-member samples.
Early works on MI use average-case success metrics, such as the attack's accuracy at guessing the membership of every sample in the evaluation set (see, e.g.,~\cite{shokri2016membership, liu2022ml}).

\Citet{carlini2022membership} critique this evaluation methodology,
noting that it does not reflect an attacker's ability
to \emph{confidently} breach the privacy of any individual sample.
They instead propose to measure the attacker's ability to infer membership---the true positive rate (TPR)---at a \emph{low false positive rate} (FPR).
Many recent works have adopted this metric
(e.g.,~\cite{chen2024hamp,bertran2023quantilemi,ye2022enhanced,zarifzadeh2024lchpmembership,wen2022coalmine}).
Yet, as we argue in this paper,
reporting the TPR and FPR aggregated over a data population
still fails to capture \emph{individual privacy}, in particular for the \emph{most vulnerable sample(s)}.
We will thus instead propose a membership inference evaluation tailored to individual least-private samples. Similar metrics to ours appear in prior work
(e.g.,~\cite{long2020pragmaticMI,carlini2022onion,jagielski2022measuring}),
but not to empirically evaluate the privacy of defenses.

\myparagraph*{DP auditing}
Differential privacy bounds an attacker's ability
to perform membership inference~\citep{kairouz2015composition}.
Specifically, for any dataset $D$ and target sample $x$,
a DP guarantee bounds the TPR-to-FPR ratio of any MI attack that distinguishes between a model trained on $D$ vs.\ $D\cup\{x\}$.
Crucially, the TPR and FPR here are calculated
with respect to the randomness of the privacy mechanism (and attacker),
but \emph{not} with respect to a random choice of the dataset $D$ or target sample $x$.
Instead, DP provides a worst-case bound on membership inference
for \emph{every choice} of dataset and target sample.

This connection can be leveraged in the opposite direction---by using
membership inference attacks to \emph{lower-bound} the DP guarantees
of an algorithm~\cite{nasr2021adversary,jagielski2020auditing,steinke2023auditing,tramer2022debugging}.
These auditing mechanisms crucially differ from typical membership inference evaluations:
to get the tightest bounds, DP auditing measures the attacker's TPR and FPR
solely for the least-private sample(s)
(often referred to as ``canaries''~\citep{carlini2019secret}),
rather than over the entire data population.

\section{Pitfalls in Privacy Evaluations}
\label{sec:pitfalls}
We identify three common pitfalls in existing empirical evaluations of privacy defenses.
As mentioned in \cref{ssec:prelims_evals},
existing evaluations typically rely on membership inference attacks,
and report some aggregate measure of attack success across a standard dataset (e.g.,~\cifar{}).
Additionally, many of these evaluations suggest that their empirical defense
achieves significantly higher utility than a differentially private baseline (e.g.,~DP-SGD).
We briefly review how existing evaluations lead to misleading empirical findings below,
and propose an evaluation protocol that more accurately reflects a defense's privacy
in \cref{sec:auditing}.

\myparagraph*{Pitfall I: Aggregating attack success over a dataset.}
Existing evaluations of membership inference attacks and defenses
report privacy metrics that are aggregated over all samples in a dataset,
either explicitly or implicitly.

Early evaluations (e.g.,~\cite{shokri2016membership, yeom2018privacy, nasr2019comprehensive})
explicitly report average metrics
such as attack accuracy or AU-ROC over a dataset of members and non-members.
These metrics thus express the average leakage of a defense across the population.
\citet{carlini2022membership} highlight a critical
issue of such metrics: they fail to
characterize whether an attacker can \emph{confidently}
infer membership of any sample
(rather than, say, just guess better than random on average).
\Citet{carlini2022membership} thus propose to
evaluate an attack's true positive rate at a low false positive rate (e.g.,~$0.1\%$),
that is, the fraction of members that the attack can identify
while making only few errors on non-members.

Yet, we note that their evaluation methodology still computes an attack's success at
identifying membership (i.e.,~the TPR) \emph{across all members}.
That is, an attacker issues guesses \emph{for all samples in the population}, and privacy leakage corresponds to the \emph{proportion} of all training set members
that are correctly identified
(while controlling the rate of false positives over the entire data population).
Informally, this evaluation thus captures \emph{how many} records in a training set can be identified while keeping the number of false guesses over the population low.

We argue that this metric (and prior ones) fail to properly capture \emph{individual privacy}.
Indeed, existing metrics view privacy leakage as a
property of a \emph{data population}, rather than  of each individual sample (i.e., does the model leak \emph{my data}?).
If a model violates the privacy of an individual,
that individual likely does not care whether the model also leaks $0.1\%$ or $10\%$
of the remaining samples;
the individual cares about the fact that an attacker can confidently recover \emph{their} data.
To make this point more concrete,
we note that existing metrics can be arbitrarily ``diluted''
by adding new members for which a defense preserves privacy,
even if the same defense fully leaks the membership of a fixed number of samples.
We illustrate this point further in \cref{sec:auditing},
where we showcase \emph{a defense that fully violates one user's privacy,
yet passes existing evaluations}.

\myparagraph*{Pitfall II: Weak or non-adaptive attacks}
Empirical defense evaluations aim to capture the privacy leakage under a realistic adversary.
It is thus important that evaluations consider strong attacks
which exploit all the capabilities of a presumed attacker.
In particular, attacks must be \emph{adaptive},
that is, fully know the defense mechanism,
and adjust their attack strategy accordingly.

Yet, in practice, many empirical defense evaluations
either use weak attacks that are no longer state-of-the-art,
or fail to adapt the attacks to peculiarities of the defense.
This situation is reminiscent of challenges in the field of adversarial examples,
where early defense evaluations misleadingly suggest robustness using non-adaptive attacks
(e.g.,~\cite{athalye2018obfuscated,tramer2020adaptive}).
For ML privacy, \citet{choquette2021label} already show
that some defenses explicitly or implicitly perturb a model's loss
to make standard membership inference attacks fail,
while remaining susceptible to different strategies.
Yet, we find that the issue of weak and non-adaptive attacks
still prevails among a number of empirical privacy evaluations.

\myparagraph*{Pitfall III: Comparison to weak DP baselines}
Given that privacy defenses with theoretical guarantees exist (e.g.,~DP-SGD),
a heuristic defense should demonstrate some clear advantage over them.
Most existing works hence argue that their proposed defense
provides a better empirical privacy-utility tradeoff\footnote{
A defense could also aim to be more computationally efficient than DP-SGD,
but few empirical defenses claim this as a main goal.
Moreover, we find that the computational cost of heuristic DP-SGD baselines is close
to the most efficient defenses we study.}
than DP-SGD---usually in the form
of higher utility at reasonable privacy.

However, we find that privacy evaluations typically consider DP-SGD baselines
that are incomparable to the proposed defense,
since the DP baselines attain only a very low accuracy.
For example, among the five evaluations in our case studies,
none considers a DP-SGD baseline with more than 80\% \cifar{} test accuracy.

The pitfall here is twofold:
First, most defense evaluations only compare to ``vanilla'' DP-SGD (as proposed in~\cite{abadi2016deep}),
without incorporating state-of-the-art techniques that can significantly boost utility
(e.g.,~\cite{de2022unlocking,sander2023TAN}).
Second, existing evaluations typically only compare to DP-SGD baselines
that achieve ``moderate'' provable guarantees (e.g.,~$\dpeps \leq 8$).
On datasets like \cifar{}, such guarantees are not achievable alongside high utility
with current techniques.
Yet, since empirical defenses forgo provable guarantees anyhow,
it makes sense to compare against a \emph{heuristic} DP-SGD baseline
with noise low-enough to achieve high utility.
While such a heuristic DP-SGD instantiation will not provide meaningful privacy guarantees,
it is a perfectly reasonable \emph{empirical} defense to consider.
Indeed, such heuristic DP uses are common in practice, with some deployments
achieving only very weak guarantees (say $\dpeps=50$)~\cite{desfontainesblog20211001}.

\section{Reliable Privacy Evaluation}
\label{sec:auditing}

To avoid misleading conclusions,
we propose a reliable and efficient evaluation protocol for empirical ML privacy.
Our protocol relies on three key points,
each targeting one of the aforementioned pitfalls we identify in existing evaluations.

\begin{enumerate}[leftmargin=20pt]
    \item Evaluate membership inference success (specifically TPR at low FPR)
    for \emph{the most vulnerable sample in a dataset},
    instead of an aggregate over all samples.
    To make this process computationally efficient, audit a set of \emph{canaries} whose privacy leakage approximates that of the most
    vulnerable sample.
    \item Use a state-of-the-art membership inference attack
    that is properly adapted to specifics of the defense.
    \item Compare to DP baselines (e.g.,~DP-SGD) that use state-of-the-art techniques
    and reach similar utility to the defense.
\end{enumerate}

In the remainder of this section,
we elaborate on each point, and discuss the practical implementation of our protocol.

\subsection{Focus on the Most Vulnerable Samples}

\begin{figure*}[t]
    \begin{subfigure}[t]{0.48\linewidth}
        \centering
        \includegraphics[width=\figonecolwidth]{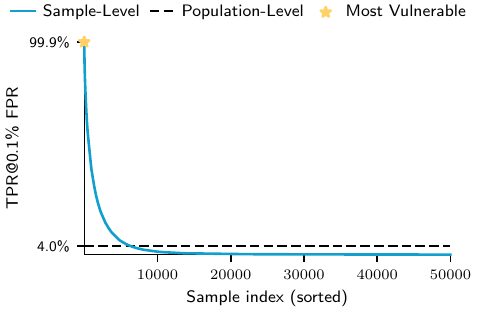}
        \caption{
        \subcaptitle{Population-level evaluations fail to capture the privacy leakage
        of the most vulnerable samples.}
        We compare the TPR@0.1\% FPR of LiRA on every \cifar{} sample in isolation (sample-level)
        to an aggregate over all samples (population-level, as in \citet{carlini2022membership}).
        The attack's success at inferring membership of the most vulnerable sample is orders of magnitude higher than the attack's success at inferring membership across the dataset.}
        \label{fig:per_sample_tpr}
    \end{subfigure}
    \hfill
    \begin{subfigure}[t]{0.48\linewidth}
        \centering
        \includegraphics[width=\figonecolwidth]{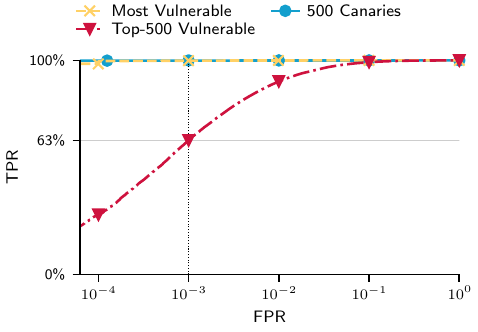}
        \caption{
        \subcaptitle{Well-chosen canaries are crucial to capture privacy of the most vulnerable sample.}
        Measuring success over the top-500 most vulnerable samples (red)
        severly underestimates the worst-case privacy leakage (yellow)
        because the sample-level TPRs in (a) decay rapidly.
        However, our well-chosen canaries
        (blue, mislabeled samples in this instance)
        closely approximate the most vulnerable sample in terms of attack success.
        }
        \label{fig:roc_common_max_ours}
    \end{subfigure}
    \vspace{-0.75em}
    \caption{
        \captitle{Membership inference evaluations should focus on the leakage
        of the most vulnerable sample(s), which can be approximated efficiently using a canary set.}
        In (a), we train 20{,}000 shadow models on \cifar{}
        to compute the MI attack success (TPR at 0.1\% FPR)
        \emph{independently for each individual sample}.
        We find that the most vulnerable sample is considerably easier to attack
        than a population-level evaluation suggests.
        In (b), we show that constructing an appropriate canary set
        allows us to capture the worst-case privacy leakage
        in a computationally efficient manner.
        Note that both plots use a linear y-axis;
        see \crefApp{app:details_for_per_sample_tpr} for experimental details.
    }
    \label{fig:average_vs_sample}
\end{figure*}

Most membership inference evaluations split a benchmark dataset $D$ (e.g.,~\cifar{})
into two disjoint sets of members $D_{\textrm{in}}$ and non-members $D_{\textrm{out}}$
(typically of equal size), and apply a membership inference attack $\mathcal{A}$.
Given a model $f$ trained on $D_{\textrm{in}}$
and a target sample $x \in D$ as input,
the attack outputs a membership score $s \gets \mathcal{A}(f, x)$
that indicates the attacker's confidence of $x$ being a member
of the training set.
Crucially, existing evaluations then quantify privacy leakage
via the aggregated attack success \emph{on every sample in $D$}---for example,
by computing an ROC curve over the attacker's confidence on all samples.
Such evaluations thus measure the \emph{fraction of samples} an attacker
can confidently identify.

We argue that this measure fails to capture
the privacy of each individual sample.
In particular, for existing evaluations,
a model's purported privacy can be arbitrarily \emph{improved} by
adding ``safe'' examples to the dataset, even though the leakage
of the most vulnerable samples does not change.
We illustrate this phenomenon with an extreme example below.

\myparagraph*{Name and shame: A blatantly non-private defense that passes existing privacy evaluations}
Consider the following simple ``name-and-shame''\footnote{
Such mechanisms often appear in discussions of
$(\dpeps, \dpdelta)$-differential privacy~\cite{smith2020lecturenotes}
to motivate the need for the parameter $\delta$ to be much smaller than
the size of the dataset.
} defense $NS_{\hat{x}}$ that fully leaks the membership of one fixed target sample $\hat{x}$:
\begin{equation*}
    NS_{\hat{x}}(D_{\textrm{in}}) = \begin{cases*}
        1 & if $\hat{x} \in D_{\textrm{in}}$, \\
        0 & otherwise.
    \end{cases*}
\end{equation*}
That is, the defense $NS_{\hat{x}}$ outputs $1$ if and only if the target $\hat{x}$
is in the training set
(obviously, this does not yield a useful ML model).
This defense completely violates the membership privacy of $\hat{x}$,
while fully protecting all other training samples.

If we evaluate any MI attack over the entire dataset $D$ of members and non-members,
we get that
\begin{equation*}
    \text{TPR} \leq \text{FPR} + 1/\abs{D} \,.
\end{equation*}
Thus, in existing evaluations,
this defense can be made arbitrarily private by increasing the size of $D$.
Evaluating on more samples indeed improves the defense's privacy \emph{on average} across the population
(i.e., a smaller proportion of the dataset risks privacy leakage);
however, from the individual perspective of $\hat{x}$,
the defense never provides any privacy at all.

\myparagraph*{Privacy is non-uniform in practice}
The name-and-shame defense is pathological,
but illustrates an important point:
the \emph{proportion} of samples whose privacy is violated
does not reflect the privacy leakage of the \emph{most vulnerable samples}.
We now show that this issue also affects privacy measurements on typical ML datasets.

We consider the standard setting from~\citet{carlini2022membership}
where the victim model $f$ is trained on a dataset $D_{\textrm{in}}$
containing half of the samples from the \cifar{} training set
(i.e., $25{,}000$ training points).
We run the LiRA attack $\mathcal{A}$ with 64 shadow models trained on random splits of \cifar{},
and evaluate the results in two ways:

\softparagraph{(1) Population-level:}
We apply the attack to each of the $50{,}000$ samples in the \cifar{} dataset
$D = D_{\textrm{in}} \cup D_{\textrm{out}}$,
and report the TPR at 0.1\% FPR across all samples---the
original setting of \citet{carlini2022membership}.
More precisely, we define
\begin{equation*}
    \tpr(\mithresh) \coloneq \, \, \Pr_{\mathclap{
    \substack{x \sim  D_{\textrm{in}} \\ f \sim \text{Train}(D_{\textrm{in}})}
    }}{\,[\mathcal{A}(f, x) > \mithresh]},
    \quad
    \fpr(\mithresh) \coloneq \, \, \Pr_{\mathclap{
    \substack{{x \sim  D_{\textrm{out}}} \\ f \sim \text{Train}(D_{\textrm{in}})}
    }}{\,[\mathcal{A}(f, x) > \mithresh]},
\end{equation*}
and select the threshold $\mithresh$ as
\begin{equation*}
    \mithresh = \argmax_\varmithresh\ \big\{\fpr(\varmithresh) \leq 0.1\%\big\} \,.
\end{equation*}
In words, we set the threshold so that the attack
makes false membership guesses for at most 0.1\% of the non-members,
and then report the \emph{proportion} of all members that are correctly identified.

\softparagraph{(2) Sample-level:}
We compute the attack's TPR and FPR for each sample \emph{individually}.
To do this, we perform the MI experiment $20{,}000$ times
by repeatedly resampling half of the \cifar{} training set,
and fitting a model on the resulting $D_{\textrm{in}}$.\footnote{
We thank Matthew Jagielski
for providing us with these models.
}
For each sample $x$ in the full \cifar{} training data,
we then define
\begin{align*}
    \tpr_x(\mithresh_x) \coloneq \Pr_{D_{\textrm{in}}, \, f \sim \text{Train}(D_{\textrm{in}})}{[
        \mathcal{A}(f, x) > \mithresh_x \mid x \in D_{\textrm{in}}
    ]}, \\
    \fpr_x(\mithresh_x) \coloneq \Pr_{D_{\textrm{in}}, \, f \sim \text{Train}(D_{\textrm{in}})}{[
        \mathcal{A}(f, x) > \mithresh_x \mid x \notin D_{\textrm{in}}
    ]},
\end{align*}
and select sample-specific attack thresholds $\mithresh_x$ as
\begin{equation*}
    \mithresh_x = \argmax_\varmithresh\ \big\{\fpr_x(\varmithresh) \leq 0.1\%\big\} \,.
\end{equation*}

That is, we now set the (sample-specific) attack threshold so that the attack
makes at most 0.1\% of false membership guesses for that specific sample,
\emph{across multiple possible training runs}.
Then, we report the probability of the attacker correctly inferring membership (again, taken over multiple training runs).

In \cref{fig:per_sample_tpr},
we rank all \cifar{} samples by their individual TPR at 0.1\% FPR (sample-level),
and compare those values to the TPR at 0.1\% FPR when aggregating attack success
across all the full dataset (population-level).
The results confirm that, even in \cifar{}, a small fraction of samples
is significantly more vulnerable to membership inference than the average data point.
In particular, the TPR at 0.1\% FPR of the most vulnerable sample is 99.9\%---orders of magnitude
higher than the population-level metric (4\%) suggests.

\myparagraph*{Our proposed evaluation metric: TPR at low FPR for the most vulnerable sample}
We thus argue that empirical evaluations of privacy defenses
should target membership inference attacks at the most vulnerable sample in a dataset,\footnote{
We could also consider a fully adversarial dataset and target sample,
as often done for auditing DP implementations~\cite{tramer2022debugging,nasr2021adversary}.
Yet, the rationale for heuristic defenses is precisely that such
worst-case scenarios are unrealistic in practice.
In keeping with this motivation,
we thus aim to measure the privacy that a defense confers for ``natural'' datasets and samples,
but while focusing on the most vulnerable of these samples.}
and report the corresponding TPR at a low FPR.
Our metric answers how likely a real-world attacker is to confidently identify
\emph{a specific sample} in the dataset, instead of measuring leakage across the entire population.

This metric reconciles empirical MI evaluations with the privacy semantics of DP,
which guard against reliable membership inference for \emph{any} individual sample,
regardless of how private a defense may be on other samples.
Our approach also accurately captures privacy of the ``name and shame'' defense:
for the most vulnerable sample, our metric yields a TPR of 100\% at 0\% FPR;
thus, the ``name and shame'' defense clearly fails to pass our evaluation.

\myparagraph*{Efficient approximation using canaries}
Ideally, our privacy evaluation would directly estimate the TPR at a low FPR
for the most vulnerable sample(s) in a dataset
by repeating the membership inference attack many times.
However, this is computationally highly expensive:
to estimate an attack's TPR at FPR $\varfpr$, even for a single sample,
we need to run the attack $O(1 / \varfpr)$ times.

If we train $\numShadow$ models and evaluate the attack on $\numCanaries$ samples,
we thus want $\numShadow \cdot \numCanaries \gg 1/\varfpr$.
This introduces a tradeoff between the tightness of the privacy bound and computational efficiency.

Existing works mainly focus on two extremes:
\begin{itemize}[leftmargin=15pt]
    \item Standard MI evaluations run the attack on the full dataset $D$
    (i.e., $\numCanaries=\abs{D}$).
    Hence, even a small number of victim models $\numShadow$ (as few as one)
    can provide sufficient statistical power if the dataset is large.
    Yet, as previously discussed, this approach yields a population-level measure of privacy.
    \item Techniques for DP auditing~\cite{nasr2021adversary, tramer2022debugging}
    often evaluate the attack on a single worst-case sample (i.e., $\numCanaries=1$),
    and must thus train $\numShadow \gg 1/\varfpr$ models
    for a strong privacy bound.
    At a typical FPR of $\varfpr = 0.1\%$, this corresponds to training thousands of models.
\end{itemize}

Our (illustrative) approach in \cref{fig:per_sample_tpr} follows the latter extreme:
we used 20,000 models to tightly approximate the per-sample attack success at low FPRs.
This approach is generally impractical,
especially since many privacy defenses add computational overhead.

We thus adopt a natural middle ground:
we evaluate membership inference on a small set of $1 \ll \numCanaries \ll |D|$ samples
(called ``canaries''~\cite{carlini2019secret}),
where each canary is inserted independently at random
in the training data of a small number $\numShadow$ of models.
The evaluation then reports membership inference success only over the canary set,
ignoring the remaining data.
Our approach resembles recent efforts to parallelize DP auditing~\cite{pillutla2023lidp}.

Crucially, we design canaries to mimic the most vulnerable samples in the data,
instead of simply selecting a subset of the data
(either at random, or in decreasing order of vulnerability).
This ensures that an attack's performance on canaries
approximates (or upper-bounds) the performance on the most vulnerable sample.
\Cref{fig:roc_common_max_ours} highlights the importance of properly designed canaries.
Here, we choose a set of $\numCanaries=500$ canaries and train $\numShadow=64$ models,
which allows us to reliably measure FPRs on the order of $0.1\%$ over the canary set.
If we were to simply pick 500 samples from the dataset at random,
we would obtain a TPR at 0.1\% FPR
close to that computed over the full dataset (i.e., $\approx 4\%$).
One might hence be tempted to use the 500 most vulnerable samples.
However, due to the small number of highly vulnerable samples,
this approach still underestimates the TPR@0.1\% FPR of the least-private sample
(63\% vs. 99.9\%).
If we instead design an appropriate canary set
(random mislabeled samples from \cifar{} in this case),
we can closely approximate the ROC curve of the most vulnerable sample---but crucially,
only train $64$ models instead of $20{,}000$
(as in \cref{fig:per_sample_tpr}).

Our approach is similar to the DP auditing procedure proposed by \citet{steinke2023auditing}.
While they focus on ``extreme'' computational efficiency by auditing an algorithm
using just one single training run (i.e., $\numShadow=1$),
we repeat the training algorithm multiple times to obtain tighter empirical privacy estimates using the LiRA attack.
As we discuss in the following,
the tightness of the estimate ultimately hinges on an appropriate choice of the
membership inference attack and canary set---both depending on the specifics of the defense.

\subsection{Adapt Attacks and Canaries to the Defense}
\label{ssec:auditing_adapt}
Reliable defense evaluations must adapt their attacks and canary choice
to the defense.
Indeed, a robust defense should protect the most vulnerable samples
against the strongest adversary within its threat model.
Yet, the nature of the most vulnerable samples, and of the strongest attack, may depend on specifics of the defense.

\myparagraph*{Adapting attacks}
State-of-the-art membership inference attacks rely on the assumption that
a model's loss (or confidence) on a sample contains the strongest membership signal.
However, some defense designs might violate this assumption.

One well-known example is \emph{confidence masking}~\cite{choquette2021label},
where a defense explicitly obfuscates
model predictions at deployment-time.
\Citet{choquette2021label} show that such defenses
are vulnerable to adaptive label-only attacks,
that is, attacks that only rely on a model's predicted label
(which those defenses preserve).

More broadly, generic membership inference attacks may be inadequate
for defenses that depart the standard supervised training regime.
For example, consider a defense based on self-supervised learning
that first trains an encoder using unlabeled data,
followed by a simple supervised fine-tuning stage.
For such a defense, memorization could occur in either of the two training stages,
but a generic attack such as LiRA might fail
to fully exploit memorization of unlabeled data during pretraining.
Similar concerns might arise for other multi-stage defenses,
for example, ones that use synthetic data generation or distillation.

\myparagraph*{Adapting canaries}
Recall that the purpose of canaries in our evaluation protocol is
to construct a set of samples such that
(1) the privacy leakage for the population of canaries approximates the leakage of the most vulnerable sample in the dataset, and
(2) the set is large enough to obtain a robust measure
of low attack FPRs (e.g.,~0.1\%) with a reasonable number of models.

\begin{figure}[t]
    \centering
    \includegraphics[width=\figonecolwidth]{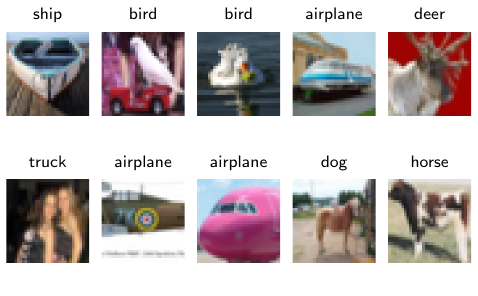}
    \vspace{-2em}
    \caption{
    \captitle{Mislabeled, ambiguous, and atypical samples are most vulnerable to privacy attacks.}
    The most vulnerable \cifar{}
    samples in the setting of \cref{fig:per_sample_tpr}
    are images that are mislabeled (e.g., humans labeled ``truck''),
    ambiguous (e.g., a bird on a car),
    or atypical (e.g., a boat on land or an airplane without wings).
    See \crefApp{app:aux_most_vulnerable} for more samples
    and TPR@0.1\% FPR values.
    }
    \label{fig:most_vulnerable_examples}
\end{figure}

Which samples are particularly vulnerable
typically depends on the type of defense that is employed.
Similarly, a defense might affect the interactions between different canaries
that are simultaneously present in the training data.
As a result, a good choice of canary is inherently defense-dependent.
\Cref{fig:most_vulnerable_examples} highlights some of the most vulnerable \cifar{} samples
for standard training,
that is, the samples with the highest TPR@0.1\% FPR in \cref{fig:per_sample_tpr}.
These examples suggest that atypical images (e.g., a ship on land) and
mislabeled samples (e.g., humans labeled as ``truck'')
are strong canary candidates for \cifar{}---at least for undefended models.

However, we must also consider interactions between canaries.
For example, suppose we use many pictures of ``boats on land''
as canaries.
Those will no longer be ``atypical'',
as the model will learn to generalize to such images.
As a result, each individual canary might exhibit significantly less memorization
than the most vulnerable sample in the original \cifar{} dataset.
We thus want canaries whose privacy leakages are as independent as possible from one another.
That is, the inclusion of one canary
in the dataset should minimally influence the
model's ability to fit the other canaries.

Mislabeled samples are a strong default candidates for canaries;
indeed, such samples are common in DP auditing~\cite{steinke2023auditing,nasr2021adversary}.
If we choose the incorrect labels at random,
a model provably cannot generalize to the canary set.
As long as the canary set is reasonably small,
using mislabeled samples also minimally affects a model's utility.
Hence, for standard supervised learning,
mislabeled samples satisfy both desiderata for good canaries,
and approximate the most vulnerable sample in the original dataset well.

Yet, similar to attacks, the choice of canaries
crucially depends on both the dataset \emph{and the defense}.
For example, as we discuss in \cref{ssec:experiments_canaries},
defenses that ignore label information are not vulnerable to mislabeled samples.
Hence, a robust empirical defense evaluation must adapt its canaries to the defense.
Doing so requires a careful analysis of the defense mechanism,
either analytically or through a form of red teaming;
systematically determining strong canaries for any given defense
is still an open problem.
We hence illustrate a heuristic approach using our case studies.

\subsection{Use Strong DP Baselines}
Since defenses with provable DP guarantees exist,
heuristic defenses should provide some distinct practical advantage over them.
In this paper, we focus on the presumed utility
advantage: heuristic defenses claim to provide higher accuracy than DP algorithms, while still
defending against realistic adversaries.

Forgoing theoretical guarantees for practical advantages is common in computer security.
For example, empirical defenses against adversarial examples
(such as adversarial training~\cite{madry2018AT}) are often preferred
over techniques with provable robustness,
as the former yield higher accuracy models while still defending against all known attacks.
More broadly, many practical deployments of cryptographic algorithms
rely on techniques with no theoretical guarantees
(e.g., hash functions like SHA-3, or symmetric encryption like AES)---even though
there exist (much more expensive) schemes whose security can be provably
reduced to a well-characterized mathematical problem (e.g.,~factorization).

While the first two principles of our methodology
focus on properly evaluating heuristic defenses,
our third principle calls for a more rigorous comparison with provable baselines.
In particular, empirical privacy evaluations should compare
heuristic defenses to state-of-the-art DP baselines
at a comparable utility level.

\myparagraph*{Focus on high utility regime}
Existing defense evaluations typically use
a DP-SGD baseline with low utility and moderately strong provable privacy guarantees
(e.g.,~$\dpeps$ often around 4--8).

Yet, a defense with a drastic utility cost is unlikely to be used in practice.
We hence argue that a comparison to low-utility DP baselines is unwarranted.
Moreover, since heuristic defenses forgo theoretical guarantees anyhow
(under the assumption that these guarantees are loose in practice),
there is no reason to hold DP baselines to the higher standard
of proper provable guarantees.

We thus instead propose to tune DP baselines such that they
(1) attain some minimal utility, comparable to that of the heuristic defense,
and (2) maximize empirical privacy under this utility constraint.
Notably, we do \emph{not enforce meaningful provable guarantees},
and potentially treat DP-SGD as a purely empirical defense.

\myparagraph*{Use state-of-the-art DP-SGD methods}
The ``vanilla'' DP-SGD algorithm of \citet{abadi2016deep}
employs a similar training setting as standard supervised learning models,
with the addition of gradient clipping and noising.
Many works show that certain techniques can substantially improve the utility of DP-SGD
while retaining the same privacy guarantees~\cite{de2022unlocking,sander2023TAN}
(e.g., by using a different data augmentation strategy).
A fair evaluation should thus account for these state-of-the-art
methods when comparing to DP-SGD.

Overall, we note that the current literature rarely studies
DP-SGD in the high-utility regime (e.g., $> 91\%$ accuracy on \cifar{}).
One potential reason is that achieving very high utility currently requires
hyperparameters (e.g., batch size and noise magnitude)
that yield meaningless worst-case guarantees (e.g., $\dpeps > 10^8$).
Yet, even without provable privacy,
DP-SGD constitutes a perfectly valid heuristic defense---which
we find to consistently outperform other methods in our case studies.

\section{Case Study Experiments}
\label{sec:experiments}
We now illustrate pitfalls in defense evaluations and motivate our evaluation strategy among five diverse empirical defenses against membership inference.
We first briefly introduce these defenses and our experimental setup
in \cref{ssec:experiments_defenses,ssec:experiments_setup}, respectively.
We then instantiate the three prongs of our evaluation strategy:
(1) we develop strong adaptive attacks in \cref{ssec:experiments_attacks};
(2) design strong canaries for each defense in \cref{ssec:experiments_canaries};
and (3) compare these empirical defenses with DP-SGD in \cref{ssec:experiments_dpsgd}
(and show that none are competitive with a properly tuned DP-SGD baseline).

We do not claim that the attacks or canaries we design for each defense are optimal,
but they suffice to highlight the stark differences
between our proposed focus on the most vulnerable sample
compared to weaker, non-adaptive population-level evaluations.
\subsection{Defenses}
\label{ssec:experiments_defenses}
The five defenses we study fall into two categories:
four are peer-reviewed defenses that explicitly aim to protect privacy,
and one is a ``folklore''
defense that illustrates how departing from a standard supervised learning setting can impede state-of-the-art attacks.
We omit certain well-known empirical membership inference defenses \cite{nasr2018machine, jia2019memguard, dong2022privacy, yang2020defending} that have been circumvented in prior work~\cite{choquette2021label, carlini2022no}.

\myparagraph*{HAMP}
HAMP~\cite{chen2024hamp} combines training-time modifications
(i.e., entropy regularization and label smoothing; not important for our attacks)
and a test-time defense that explicitly randomizes a model's confidence.
Specifically, given a trained model $f$ and input image $x$,
the defense output a \emph{random} confidence vector
such that the order of predicted classes matches the original prediction $f(x)$.
This is an obvious case of \emph{confidence masking}~\cite{choquette2021label}.

\myparagraph*{RelaxLoss}
RelaxLoss~\cite{chen2022relaxloss} reduces overfitting by constraining the training loss to be above a fixed threshold.
Concretely, for every training batch,
the defense first computes the cross-entropy loss
$\ell_{\text{batch}} = 1/B \sum_{i=1}^B{\mathcal{L}(x_i, y_i)}$ as in standard training.
Then, RelaxLoss compares the batch loss to a target loss threshold $\rlthresh$:
If $\ell_{\text{batch}} > \rlthresh$, the defense continues with standard gradient descent;
however, if $\ell_{\text{batch}} \leq \rlthresh$ then RelaxLoss instead takes
a (modified) gradient \emph{ascent} step, with the goal of raising the loss above $\rlthresh$.

\myparagraph*{Self Ensemble Architecture (SELENA)}
SELENA~\cite{tang2022selena} is a distillation defense
that heuristically mimics the provable guarantees of PATE~\cite{papernot2016semi},
without the need for public data or noise addition.
SELENA first splits the training data into (partially overlapping) chunks $D_1, D_2, \dots, D_k$
and independently trains one teacher model $f_i$ on each chunk.
In a second distillation stage, the defense trains a model $f_{\text{student}}$
using the soft predictions from the $k$ teacher models.
To promote membership privacy, for every sample $x$,
SELENA only distills soft prediction from the teachers $f_i$
that were \emph{not} trained on $x$ (i.e., $x \notin D_i$).
The rationale is that $f_{\text{student}}$ is trained to mimic the responses
of the teachers on non-members only.

\myparagraph*{Data-Free Knowledge Distillation (DFKD)}
DFKD~\cite{lopes2017data,yin2020dreaming, chen2019data,zhang2023ideal}
transfers knowledge from a teacher model---trained on private data---to a student model
trained solely using \emph{synthetic data}.
While data privacy is a primary motivation for DFKD,
we are not aware of prior work evaluating this defense
against membership inference attacks
(some works argue privacy
by visually comparing the synthetic data to the training data~\citep{hao2021data,zhang2022dense}).
As a representative from this line of work,
we study the state-of-the-art method of~\citet{fang2022up}.
At a high level, their method proceeds in four steps:
(1) train a teacher model $f$ on the private training set;
(2) train a generative model to produce synthetic data using an inversion loss
(e.g., by matching the batch-normalization statistics of the teacher model);
(3) distill a student model $f_{\text{student}}$ using synthetic images from the generator
and soft labels from the teacher model;
(4) repeat steps 2 and 3 iteratively until the model converges.
The privacy intuition is that the student model only observes noisy synthetic data,
not the original (private) training data.
Yet, as we will see, such ``visual'' privacy arguments can be highly misleading.

\myparagraph*{Self-Supervised Learning (SSL)}
Self-Supervised Learning is a technique
to learn feature representations from unlabeled data.
Given a labeled dataset $\{(x_1, y_1), \dots, (x_n, y_n)\}$,
the SSL defense first trains a feature encoder
$\phi$ in an unsupervised fashion, using only the features $\{x_1, \dots, x_n\}$.
We consider two popular methods, SimCLR~\cite{simclr} and MoCo~\cite{moco},
both employing a contrastive loss which ensures that
different \emph{augmentations} of an input yield similar features
(i.e., $\phi(x) \approx \phi(\texttt{aug}(x))$).
Then, in a second stage, we train a linear classifier
$f(x) = W \phi(x) + b$ on top of the frozen encoder,
using the full labeled training set and a standard cross-entropy loss.
SSL is not explicitly a defense against membership inference,
but has received a lot of study regarding its privacy~\cite{liu2021encodermi, he2021quantifying, ko2023practical}.
We include this SSL-based defense to illustrate a shortcoming of a naive privacy evaluation
that applies a LiRA-like attack out-of-the-box
without accounting for the unsupervised nature of the encoder.

\subsection{Experimental Setup}
\label{ssec:experiments_setup}

\myparagraph*{Dataset}
We run all experiments on \cifar{}~\cite{krizhevsky2009learning},
a canonical benchmark dataset used by most existing empirical evaluations of privacy defenses.
Due to the relatively high computational cost of our experiments,
we refrain from studying more datasets and focus our efforts on a single one.
As our goal is to reveal pitfalls in existing evaluations,
we believe that case studies on the most standard dataset used in the field are sufficient.
Similarly, previous works on pitfalls in adversarial robustness evaluations
typically use a single dataset to show that existing evaluations
are incomplete~\cite{athalye2018obfuscated, tramer2020adaptive}.

\myparagraph*{Shadow models and audit samples}
Similar to \citet{carlini2022membership}, we train multiple models
on random subsets of the \cifar{} training set.
However, rather than subsampling the entire training set as in \cite{carlini2022membership},
we follow \citet{steinke2023auditing}:
we designate 500 random data points
as ``audit samples'' on which we evaluate membership inference;
we always include the remaining 49{,}500 samples in every model's training data.\footnote{
Fixing most of the training data may put adversaries at an advantage,
as the shadow models and target models are more similar in our setting.
Yet, we show in \crefApp{app:aux_validate_loo} that this has a negligible effect
on attack success rates.}
For population-level evaluations, we attack these audit samples as-is
(since the audit samples are a random sample from the population,
the expected attack success on the audit samples is the same as on the full population).
For our evaluation that focuses on the most vulnerable samples,
we replace the 500 audit samples with appropriately chosen canaries.

For each defense, we train 64 models,
randomly including each audit sample in exactly half of the models' training datasets.
For evaluation, we use a leave-one-out cross-validation (as in \cite{tramer2022truthserum}),
where we evaluate the attack 64 times,
once with each model as the victim and the remaining 63 models as the attacker's shadow models.
We then calculate the attack's TPR and FPR over the $64 \cdot 500$ guesses
of the attacker on all canaries and victim models.

To control for randomness in our evaluation,
all experiments use the same non-audit samples,
shadow model assignments,
and audit samples given a fixed choice of canaries.
Hence, two experiments using the same type of canaries use exactly the same data;
datasets with different canaries are identical up to the 500 audit samples.

\myparagraph*{Defense implementation}
We further control neural network architecture and capacity:
all defenses except SSL use a WRN16-4 base model~\cite{zagoruyko2017wideresnet};
for SSL, we could not achieve sufficiently high utility with the WRN16-4 architecture,
and thus follow \cite{moco,simclr} by using a ResNet-18~\cite{he2016resnet} instead.

Moreover, we re-implement all defenses (carefully following all original design decisions).
This allows us to use exactly the same setting in all case studies,
and enables straightforward reproducibility of our results.
We then tune all privacy-related defense hyperparameters (where available) to maximize privacy
constrained to at least 88\% \cifar{} test accuracy.
and otherwise use the values proposed in each defense's original paper.
See \crefApp{app:hps} for specific hyperparameters and implementation details.

\myparagraph*{LiRA attack}
For LiRA, we always report the maximum TPR@0.1\% FPR
over the strongest approaches proposed in~\cite{carlini2022membership}.
More precisely, we consider the Hinge vs.\ Logit scores,
and attacking just the original sample vs.\ 18 augmented versions
(since not all defenses employ data augmentation).
The augmentations consist of horizontal flips,
and shifting images by $\pm 4$ pixels on each axis.

\subsection{Adaptive Attacks}
\label{ssec:experiments_attacks}
\begin{figure*}[t]
    \begin{subfigure}[t]{0.48\linewidth}
        \centering
        \includegraphics[width=\figonecolwidth]{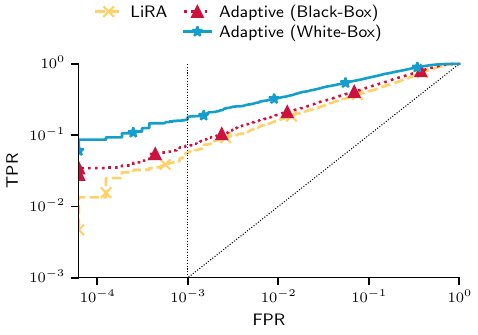}
        \caption{
        \subcaptitle{Adapting to the pretraining loss yields a stronger attack for SSL.}
        In both a black-box (attacking the final classifier)
        and especially in the white-box setting (directly attacking the SSL encoder),
        our score that exploits contrastive losses
        increases the success of the standard LiRA attack.
        We only present the SimCLR-based defense for brevity;
        see~\crefApp{app:aux_ssl} for MoCo.
        }
        \label{fig:ssl_attack_simclr}
    \end{subfigure}
    \hfill
    \begin{subfigure}[t]{0.48\linewidth}
        \centering
        \includegraphics[width=\figonecolwidth]{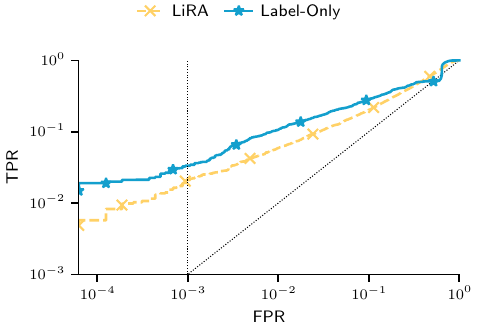}
        \caption{
        \subcaptitle{A label-only attack circumvents confidence masking in HAMP.}
        The test-time defense provides good privacy against
        state-of-the-art confidence-based attacks such as LiRA (yellow),
        but our simple label-only attack (blue) achieves almost threefold privacy leakage.
        See \crefApp{app:aux_hamp} for more details.
        }
        \label{fig:hamp_labelonly}
    \end{subfigure}
    \vspace{-0.75em}
    \caption{
        \captitle{Adaptive membership inference attacks that exploit defense-specific mechanisms
        improve over the standard LiRA attack.}
        We show results for the (a) SSL and (b) HAMP defenses, with LiRA evaluated across the entire dataset.
    }
\end{figure*}

A reliable privacy evaluation must use the strongest possible attack in a given threat model.
We hypothesize that for two defenses in our case studies---SSL and HAMP---the standard LiRA
attack is not strong,
because both defenses violate some of the attack's implicit assumptions.
We thus develop custom attacks tailored to the specifics of SSL and HAMP.
While we do not claim either attack to be optimal,
our simple adaptations suffice to highlight how evaluations using weak attacks
can yield misleading results.

\myparagraph*{Adapting attacks to contrastive losses in SSL}
SimCLR and MoCo, the two SSL techniques that we consider,
train an encoder neural network $\phi$ using a contrastive loss
to learn representations from \emph{unlabeled} images.
The full defense trains an additional linear classifier on
top of this (fixed) encoder, using the private labels.

We hypothesize that the cross-entropy loss of the full defense
encodes only a weak membership signal, since this loss is only used to train the
final layer. Thus, applying LiRA out-of-the-box to the full defense is unlikely
to be effective. This has also been highlighted in concurrent work~\citep{wang2024sslmemorization}, which shows that
SSL encoders tend to memorize training images despite not using any labels for training.
We hence adapt LiRA to the SSL setting by specifically targeting the contrastive loss used to train the encoder.

When training an encoder, both MoCo and SimCLR maximize the similarity between
representations of augmented version of the same image (positive pairs),
and minimize the similarity between augmented versions of different images (negative pairs).
We thus expect that representations of an image under different augmentations
are more similar if that image is a training member---analogous to overconfident predictions
in supervised learning.

Specifically, given a target image, we apply two random augmentations to the image, and calculate the cosine similarity between the corresponding defense outputs.\footnote{Since the cosine similarity is a value $\rho \in [-1, 1]$,
we apply a Fisher transformation, $\log{(1+\rho)/(1-\rho)}$,
to obtain empirically normally distributed statistics.}
We consider two attack variants here: (1) a white-box attack that directly computes similarity over the outputs of the \emph{encoder} $\phi$; (2) a black-box attack that applies the contrastive loss to the logits output by the full defense (including the linear classifier).
Finally, to account for randomness in the data augmentations,
we repeat this procedure six times for every image, and average the similarities.

The results in \cref{fig:ssl_attack_simclr} (for SimCLR) support our hypothesis
that most memorization happens during self-supervised training.
Directly attacking the SSL encoder (white-box) using our adaptive attack
yields a threefold increase in privacy leakage compared to standard LiRA
on the full defense.
In a black-box setting, our adaptation of the attack's loss increases the TPR against the full defense from 5.8\% to 7.1\% (at a FPR of 0.1\%).
In the remainder of this section, we build upon the stronger white-box attack,
and present black-box results in \crefApp{app:aux_ssl}.

While our results
already highlight the importance of strong adaptive attacks,
more sophisticated strategies might reveal even higher privacy leakage.
Indeed, our current attack only considers the positive part of the contrastive loss (i.e., similarity between two augmentations of an image), while ignoring the negative part (i.e., dissimilarity between augmentations of different images).

An orthogonal, but interesting observation from this experiment is that there do exist defenses where a white-box MI adversary outperforms a black-box attacker. As noted by \citet{carlini2022membership}, it remains unknown whether we can build stronger white-box attacks for standard supervised learning defenses.

\myparagraph*{Circumventing confidence masking in HAMP with label-only attacks.}
In contrast to SSL, HAMP uses a fairly standard supervised learning approach, for which LiRA is appropriate.
However, since the defense actively obfuscates the model's predicted confidences at test time, the standard attacks achieve a low
TPR of only 2.1\% at 0.1\% FPR.
This (presumed) protection stems primarily from HAMP's test-time defense,
as the same attack on non-obfuscated predictions
is significantly stronger (see \crefApp{app:aux_hamp}).

The test-time defense erases all information in model predictions
besides the predicted label order,
thereby performing \emph{confidence masking}~\citep{choquette2021label}.
We hence follow \citet{choquette2021label},
and develop a straightforward label-only attack.

Our attack queries the model on 18 fixed data augmentations
of the target sample, and checks whether the model classifies each input correctly.
This yields a binary vector of 18 entries.
Using the shadow models, we then fit a logistic regression classifier,
which takes this binary vector as input, and predicts membership of the target sample.
Finally, we use the confidence of each classifier as a membership score $\mathcal{A}(f; x)$, and calculate the usual TPR and FPR statistics.
See \crefApp{app:hps} for further details.

\Cref{fig:hamp_labelonly} compares our label-only attack with standard LiRA.
Our adaptive label-only attack achieves a consistent
increase in TPRs compared to the original LiRA attack.
Based on the results in~\cite{choquette2021label},
we conjecture that computationally more expensive label-only attacks
(e.g., using strategies from black-box adversarial example attacks)
would be even stronger.
But, as we will see in \cref{ssec:experiments_canaries}, our simple adaptive attack suffices
to break the privacy of HAMP on the most vulnerable samples.

\myparagraph*{Other defenses.}
The other heuristic defenses we study (Relaxloss, SELENA and DFKD)
use more standard supervised learning methods,
without any obvious confidence masking.
We thus apply the original LiRA attack to these,
and show in the following section that this suffices to breach privacy of the most vulnerable samples.

Nevertheless, it is possible that stronger adaptive MI attacks exist for some of these defenses.
In particular, DFKD's use of generative modeling and synthetic data
introduces a layer of indirection that could be exploited.
However, our attempts at building a stronger attack than LiRA for this defense were unsuccessful.

\subsection{Sample-Level Privacy using Canaries}
\label{ssec:experiments_canaries}

We now focus on the most important part of our evaluation protocol:
measuring the attack success on the most vulnerable samples in a dataset,
rather than on the dataset as a whole.
Recall that evaluating attack success on each sample independently
would be computationally expensive, as it requires thousands of shadow models.
We instead measure the attack's success on a set of canaries
that are designed to mimic the (suspected) most vulnerable sample.

\begin{figure}[t]
    \centering
    \includegraphics[width=\figonecolwidth]{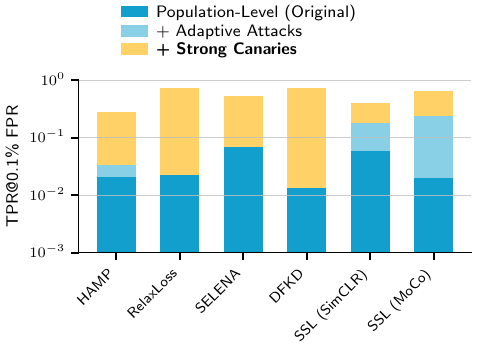}
    \vspace{-1em}
    \caption{
    \captitle{Our improved evaluation uncovers substantial privacy leakage
    for the most vulnerable samples.}
    While most defenses appear private at the population-level (original),
    our evaluation with strong adaptive attacks
    targeted at defense-specific canaries reveals large privacy leakage for the most vulnerable samples.
    \emph{Note that the defense that appears to be the most private on a population-level (DFKD)
    is the second-least private on a sample-level!}
    }
    \label{fig:avg_vs_worst}
\end{figure}

\Cref{fig:avg_vs_worst} provides a summary for all our case studies.
We consistently find that approximating the most vulnerable samples
using defense-specific canaries yields a TPR@0.1\% FPR
that is between $2\times$ and over $50\times$ higher
compared to population-level evaluations on a random \cifar{} subset.
Crucially, our evaluation substantially changes the ranking between defenses:
DFKD, for example, appears to be one of the most private defenses when success is aggregated over the full dataset,
yet exhibits the second-worst privacy leakage for the most vulnerable samples.
We discuss our canary choices (\cref{tab:canaries} in the appendix)
and stress the importance of adapting canaries to evaluated defenses
in the remainder of this section.

\myparagraph*{Mislabeled samples are a strong baseline}
As discussed in \cref{ssec:auditing_adapt},
mislabeled samples are a good default choice of canaries.
Intuitively, those samples are naturally vulnerable to membership inference
in supervised learning:
a model capable of memorization and generalization
will tend to assign high confidence to the wrong class
if a mislabeled sample is a training member (memorization),
but low confidence if the sample is not in the training data (generalization).
Because practical datasets tend to contain some label noise
(e.g.,~\cite{muller2019mislabeled,zhang2017mislabeled,northcutt2021labelerror}),
mislabeled samples may hence approximate the most vulnerable samples in such datasets well.

To generate mislabeled samples as canaries,
we independently change the labels of all 500 audit samples
to a uniformly random new class.
For HAMP, RelaxLoss, and DFKD, a MI attack on those canaries yields
a TPR@0.1\% FPR of around 30\% to 70\% in \cref{fig:avg_vs_worst}.

For DFKD in particular, this highlights the importance of rigorous privacy evaluations,
compared to visual inspections or intuitions (as done in some prior work~\cite{hao2021data,zhang2022dense}).
Even though DFKD's distillation process uses synthetic data---and hence omits wrong labels---the defense exhibits a TPR@0.1\% FPR of $72.2\%$.
We defer a more detailed investigation to future work,
and now focus on two defenses that are robust against label noise: SSL and SELENA.

\begin{figure*}[t]
    \begin{subfigure}[t]{0.48\linewidth}
        \centering
        \includegraphics[width=\figonecolwidth]{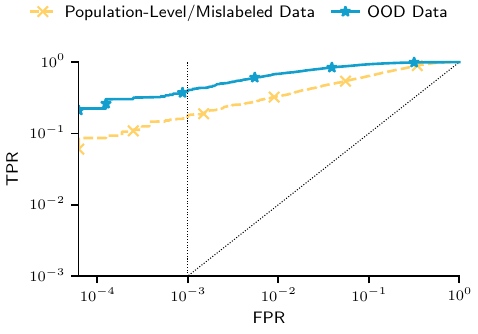}
        \caption{
        \subcaptitle{OOD samples are strong canaries for SSL defenses.}
        For SimCLR, using our adaptive white-box attack,
        OOD ImageNet data (blue)
        is more vulnerable to membership inference
        than the population average (yellow).
        Note that neither the SSL encoder nor our white-box attack depend on labels;
        hence, mislabeling an audit sample does not change its privacy leakage.
        See \crefApp{app:aux_ssl} for more details and the MoCo-based defense.
        }
        \label{fig:ssl_ood}
    \end{subfigure}
    \hfill
    \begin{subfigure}[t]{0.48\linewidth}
        \centering
        \includegraphics[width=\figonecolwidth]{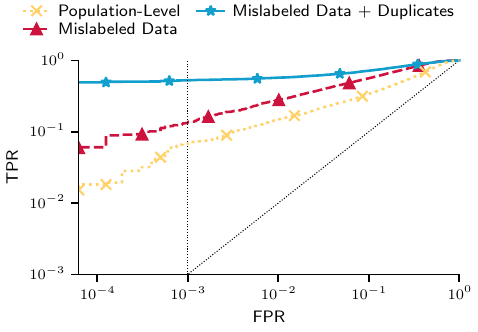}
        \caption{
        \subcaptitle{(Near)-duplicates in the training data
        create particularly vulnerable samples for SELENA.}
        Mislabeled canaries in isolation (red)
        only mildly increase privacy leakage over a population-level evaluation (yellow).
        However, those mislabeled samples become highly vulnerable
        if they have a duplicate in the training set (blue).
        }
        \label{fig:selena_student}
    \end{subfigure}
    \vspace{-0.75em}
    \caption{
        \captitle{The right choice of canary set reveals privacy leakage
        that is much higher than when the attack success is aggregated over the entire dataset.}
        Different defenses require different canaries.
        While mislabeled samples are often a reasonable choice, we find that (a) OOD data for SSL and (b) mislabeled \emph{duplicated} samples for SELENA
        are particularly vulnerable.
    }
\end{figure*}

\myparagraph*{Out-of-distribution images are strong SSL canaries}
Real-world data is often long-tailed,
that is, it contains many ``typical'' samples, and few outliers
(e.g., mislabeled or atypical images).
For standard supervised learning, \citet{feldman2020memorization} argues that
the only way to fit outliers and perform well on
similar test samples is by memorizing labels.
However, self-supervised learning itself solely relies on unlabeled data;
changing a sample's class does hence not influence memorization.
We thus hypothesize that our SSL defenses,
which heavily rely on pretraining with unlabeled data,
do not memorize most label noise.
In fact, we find that both SSL defenses have an average training accuracy on mislabeled canaries
of only 1.3\%---significantly below random guessing (10\%).

Hence, we consider \emph{atypical images} as a different type of outliers.
Indeed, recent work by~\citet{wang2024sslmemorization} found
that SSL feature encoders
tend to memorize atypical images,
and such memorization can be necessary for good downstream generalization.

However, since rare images constitute only a small fraction of \cifar{} by definition,
we use a proxy as canaries: \emph{out-of-distribution (OOD) data}.
More concretely, we replace the original audit set
with 500 downsampled ImageNet images.
To decrease correlation between canaries,
we pick each sample from a different ImageNet class,
and assign labels independently at random.

The results in \cref{fig:ssl_ood} confirm our hypothesis,
and highlight how the choice of canaries depends on the specifics of a defense.
In a white-box setting,
our attack on OOD canaries achieves a TPR@0.1\% FPR
as high as 65\%---between 2.2$\times$ and 2.7$\times$ times higher
than on the original/mislabeled audit set.

The choice of canaries is even more important in the black-box setting.
Indeed, mislabeled samples yield a slightly \emph{lower} TPR@0.1\% FPR
compared to the original (in-distribution) audit set.
In contrast,
OOD canaries are much more vulnerable.
We defer those and additional results to \crefApp{app:aux_ssl} for brevity.

\myparagraph*{(Near)-duplicates are strong canaries for SELENA}
Recall that SELENA first trains an ensemble of models $f_1, \dots, f_\saiModels$
on overlapping subsets of the training data.
SELENA then distills each sample $x$ into a student model
using only predictions from the teacher models that were \emph{not} trained on $x$.
\citet{tang2022selena} prove that SELENA's ensemble mechanism
leaks nothing about a sample's membership
if queried only on that specific sample.

Unfortunately, this proof ignores \emph{interactions} between samples.
Concretely, suppose that training a model on a mislabeled sample $(x, \hat{y})$
also affects the model's prediction on a different training image $x'$,
for example, because $x'$ and $x$ are similar.
When SELENA distills $x'$ into the student model,
it will only query teachers $f_i$ not trained on $x'$.
However, some of these teachers will have been trained on the mislabeled sample $(x, \hat{y})$.
In that case, we expect that $f_i(x') \approx f_i(x) = \hat{y}$,
that is, the incorrect label may \emph{leak} into the student via predictions on other samples.
We find evidence of this effect in practice:
if we evaluate SELENA on mislabeled canaries, we get a TPR of 13.8\% at 0.1\% FPR,
about twice as high as for the correctly labeled audit set.

Thus, SELENA might not protect some samples due to \emph{other samples with similar features}
in the training data (we investigate this more thoroughly in \crefApp{app:aux_selena}).
More precisely, we expect that the most vulnerable samples for SELENA
are mislabeled images $(x, \hat{y})$
for which a \emph{near-duplicate} $x' \! \approx x$ also exists in the training set
($x'$ may be correctly labeled).

\Citet[App.\ A.3]{tang2022selena} conjecture that such samples are unlikely to exist in practice.
Yet, we find that \cifar{} does contain mislabeled/ambiguous samples
with near-duplicates in the training data
(see the examples  %
\ifarxiv{in \cref{fig:different_labels_nns}}\else\fi
in \crefApp{app:aux_selena}).

This inspires our canary choice:
We duplicate half of the original audit set, and mislabel one sample per pair.
We then use these 500 samples as the new audit set
(i.e., we randomly include each of those samples as a member with 50\% probability)
but we only evaluate the attack on the mislabeled 250 instances.
\Cref{fig:selena_student} confirms our hypothesis:
attacks on mislabeled samples with duplicates in the audit set
achieve a TPR of 52.7\% at 0.1\% FPR---a roughly $3.8\times$ increase
compared to mislabeled samples in isolation.

Note that the TPR is just above 50\%, even at very low FPRs.
This is because the attack succeeds when the mislabeled sample $(x, \hat{y})$ is a member,
\emph{conditioned on the near-duplicate $x'$ also being a member}.
Since we vary membership of all audit samples independently at random,
this happens with probability $50\%$---bounding the expected number of successes.
In \crefApp{app:aux_selena}, we consider a stronger setting,
where the near-duplicates are always part of the training data;
we find that this enables near-perfect membership inference.

\subsection{Strong DP-SGD Baselines}
\label{ssec:experiments_dpsgd}

As we have shown, none of the heuristic defenses we study provides reasonable privacy protection
for the most vulnerable samples.
We hence ask whether this indicates that such leakage is inherent for a high-accuracy model trained on \cifar{},
or if other heuristic defenses could provide a better tradeoff.

We consider DP-SGD~\cite{abadi2016deep} as a natural baseline to answer this question.
However, for a fair comparison with other heuristic defenses,
we focus on a high-utility regime;
that is, we view DP-SGD as a purely heuristic defense
while possibly forgoing meaningful provable guarantees.
Concretely, we consider two DP-SGD instances:
a medium utility baseline that maximizes empirical privacy
constrained to at least 88\% \cifar{} test accuracy,
and a high utility baseline with 91\% test accuracy.
We show that high-utility DP-SGD yields a very competitive privacy-utility tradeoff, surpassing all the other heuristic defenses we consider (at a similar utility level).

\myparagraph*{DP-SGD baselines}
Both baselines rely on state-of-the-art DP-SGD
training techniques~\cite{de2022unlocking,sander2023TAN}.
We use a modified WRN16-4 architecture that replaces batch normalization
with group normalization, swaps the order of normalization and ReLU,
and uses the custom initialization scheme of~\cite{de2022unlocking}.
We further employ augmentation multiplicity~\cite{de2022unlocking} using the modified Opacus \cite{opacus} library of \citet{sander2023TAN},
and return an exponential moving average of the model weights with decay factor $0.9999$.

We tune the hyperparameters of the medium utility baseline in the same way
as for all case studies (see \cref{ssec:experiments_setup}), that is, to
maximize privacy (measured by $\dpeps$ at $\dpdelta = 10^{-5}$)
subject to at least $88\%$ \cifar{} test accuracy.
Crucially, we do \emph{not} enforce meaningful DP guarantees;
the provable privacy guarantees $\dpeps$ of all our DP-SGD baselines are in the thousands.
For the high utility baseline,
we rely on recent scaling laws~\citep{sander2023TAN}
to increase the medium baseline's utility at the cost of privacy
(primarily by decreasing batch size while carefully scaling noise).
See \crefApp{tb:dpsgd_params_full} for all hyperparameters
and the average \cifar{} test accuracy over 64 models.

\begin{figure}[t]
    \centering
    \includegraphics[width=\figonecolwidth]{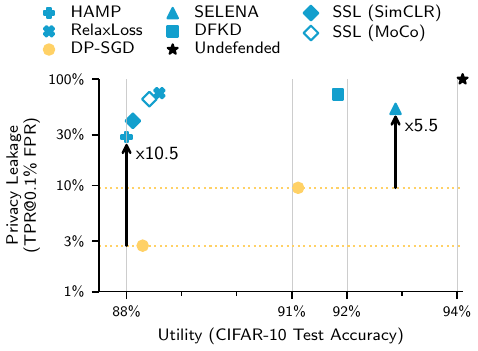}
    \caption{
    \captitle{DP-SGD is a strong empirical privacy defense.}
    We compare the privacy-utility tradeoff of the 5 heuristic defenses we study
    to two DP-SGD baselines tuned for high test accuracy.
    For both baselines, the DP-SGD privacy analysis yields near-vacuous guarantees
    ($\dpeps \gg 1{,}000$),
    yet both provide much better empirical privacy for the most vulnerable samples, according to our evaluation.
    }
    \label{fig:dpsgd_comparison}
\end{figure}

\myparagraph*{Adaptive attacks and canaries for DP-SGD}
We consider the same threat model as in our case studies,
in contrast to typical DP-SGD audits,
where adversaries can observe and influence all model updates~\cite{nasr2021adversary}.

As canaries, we consider three types of outlier data
that have been used for DP-SGD auditing in prior work~\cite{nasr2023tight,steinke2023auditing,nasr2021adversary}:
mislabeled samples, OOD data, and uniform images.
We explicitly omit adversarial examples (as used in~\cite{nasr2021adversary}),
since auditing many of these samples in parallel
induces a weak (non-adaptive) form of adversarial training~\citep{tramer2017ensemble}.
This would cause each individual canary to be less effective,
as the partially robust model would be more likely to correctly classify the canaries
that are \emph{not} in the training set.

For attacks, we consider the same setting as for other defenses,
where the attacker only gets access to the final model after training.
In the DP-SGD literature, this is often called a black-box attacker,
as opposed to a white-box attacker who can see each noisy gradient step.
In this threat model, there are (to the best of our knowledge) no known adaptive attacks on DP-SGD
that outperform standard attacks like LiRA.
We thus use LiRA on the final trained model,
but report the maximum TPR@0.1\% FPR over all canary types.

\myparagraph*{Results}
\Cref{fig:dpsgd_comparison} compares the heuristic defenses we consider to our DP-SGD baselines
(all evaluated according to our protocol, with adaptive attacks and strong canaries).
Despite meaningless provable guarantees ($\dpeps > 10^8$),
our high utility DP-SGD baseline shows decent empirical privacy:
all heuristic defenses with similar test accuracy yield a TPR@0.1\% FPR
that is at least $5.5\times$ worse.
Compared to the medium utility DP-SGD baseline ($\dpeps > 3{,}000$),
even the most private heuristic defense we study (HAMP)
leaks $10\times$ more membership privacy,
at a slightly worse test accuracy.

Two defenses in our case studies (DFKD and SELENA) achieve
slightly higher utility than our best DP-SGD baseline ($\approx 92$--$93\%$ \cifar{} test accuracy)---albeit at the cost
of much higher privacy leakage.
This raises the question if \emph{any} defense can provide meaningful membership privacy
for the most vulnerable samples in this very high utility regime
(without using public data).
There is evidence to suggest that the answer may be negative.
In particular, the work of \citet{feldman2020memorization} proves that classifiers
trained on heavy-tailed data distributions necessarily need to memorize some training labels
to achieve optimal generalization.
Correctly classifying the tail of the \cifar{} test data might thus require
memorization of similar rare examples during training,
rendering those examples susceptible to membership inference.
To give more credence to this hypothesis, \crefApp{app:aux_dpsgd} shows
that even DP-SGD fails to provide reasonable privacy
when pushed to reach around 92\% test accuracy.

Ultimately, we do not claim that DP-SGD
is the best membership inference defense
in all settings.
Yet, we show
how---even absent meaningful provable guarantees ($\dpeps \gg 1{,}000$)---DP-SGD
is a strong \emph{empirical} defense with competitive utility on \cifar{}.
Thus, future heuristic defenses that aim to claim a better privacy-utility tradeoff than DP-SGD should show a clear advantage over our baselines.

\section{Conclusion}
Throughout this paper, we have illustrated three major methodological pitfalls
in empirical privacy evaluations using membership inference attacks.
Existing evaluations report metrics that do not convey meaningful individual privacy semantics,
use weak attacks, and consider subpar DP baselines.
The evaluation methodology we propose is one way to fix these issues.
Our work exposes a number of possible takeaways and future research directions.

\myparagraph*{Privacy semantics in-between average-case and worst-case.}
As we show, the exact way we measure the privacy of a defense matters a lot.
Before evaluating a defense---or an attack---we thus need to clearly define the privacy \emph{semantics}
that the evaluation targets
(e.g., do we care about the proportion of samples that can be inferred,
or if \emph{any} sample can be inferred).

These privacy semantics are often implicit in the formal membership inference game that a work starts from
(e.g., are the dataset and target sample chosen randomly or by the adversary?),
but this is rarely explicitly discussed.
Ultimately, these choices interpolate between an
average-case setting---where the data and target are randomly chosen---and
a worst-case setting---where the dataset and target are adversarial.
The design of heuristic privacy defenses is often motivated by the fact
that the latter worst-case setting
is overly pessimistic.
But this need not imply that the other extreme
(the fully average-case setting) is appropriate either.

\myparagraph*{A theory of ``natural'' privacy leakage.}
A possibly surprising finding from our work is that ``heuristic DP-SGD''
(with hyperparameters that do not provide meaningful provable guarantees)
is by far the best defense in practice.
Yet, the DP-SGD analysis is tight
in worst-case (possibly pathological) settings~\cite{abadi2016deep, nasr2021adversary, feng2024privacy}.
A formal understanding of DP-SGD's performance in ``natural'' settings
might thus lead to tighter provable privacy under realistic assumptions.

Another intriguing question raised by our work (and others~\cite{nasr2021adversary, jagielski2020auditing})
is how to create strong canaries for a given defense.
That is, how do we design or efficiently identify samples
that are most vulnerable to privacy attacks?
In our setting, an additional goal is to design a \emph{collection} of such samples,
where each sample is independently highly vulnerable.
For now, we rely primarily on heuristics to select such samples,
rather than on a principled approach.

\myparagraph*{DP-SGD is a pragmatic defense.}
A welcome finding from our work is that DP-SGD may be the ``best-in-class'' defense to apply in practice,
whether one cares about stringent provable privacy guarantees or not.
As a result, a single infrastructure and set of tools can be used
for cases where data privacy is paramount (by setting hyperparameters to get strong provable privacy),
as well as for cases where absence of memorization is a ``nice to have'' (by setting hyperparameters to get high utility).

\begin{acks}
M.A.\ and J.Z.\ are funded by the
\grantsponsor{SNSF}{Swiss National Science Foundation (SNSF)}{https://data.snf.ch/grants/grant/214838}
project grant \grantnum{SNSF}{214838}.
We thank Matthew Jagielski for providing us with the 20,000 \cifar{} models
used in \cref{fig:average_vs_sample}.
\end{acks}

\bibliographystyle{ACM-Reference-Format}
\balance
\bibliography{main}

\appendix
\ifarxiv%
\clearpage
\begin{table*}[t]
    \caption{Full privacy (TPR@0.1\% FPR) and utility (\cifar{} test accuracy) values for each defense.}
    \label{tab:metrics}
    \begin{tabular}{@{}lrrcrr@{}}
    \toprule
    Method & \multicolumn{2}{c}{TPR@0.1\% FPR} & \phantom{abc} & \multicolumn{2}{c}{\cifar{} Test Accuracy} \\
    \cmidrule{2-3} \cmidrule{5-6}
    & Population-Level, LiRA & Sample-Level, Adaptive Attack && Population-Level & Sample-Level \\ \midrule
    HAMP & 2.1\% & 28.5\% & & 88.29\% ($\pm 0.04$) & 88.00\% ($\pm 0.04$) \\
    RelaxLoss & 2.2\% & 74.1\% & & 88.86\% ($\pm 0.03$) & 88.60\% ($\pm 0.03$) \\
    SELENA & 6.8\% & 52.7\% & & 93.05\% ($\pm 0.02$) & 92.88\% ($\pm 0.02$) \\
    SSL (SimCLR) & 5.8\% & 40.6\% & & 88.18\% ($\pm 0.02$) & 88.11\% ($\pm 0.03$) \\
    SSL (MoCo) & 2.0\% & 65.0\% & & 88.44\% ($\pm 0.04$) & 88.41\% ($\pm 0.04$) \\
    DFKD & 1.3\% & 72.2\% & & 92.39\% ($\pm 0.03$) & 91.84\% ($\pm 0.03$) \\
    DP-SGD (medium utility) & 0.7\% & 2.7\% & & 88.29\% ($\pm 0.03$) & 88.29\% ($\pm 0.03$) \\
    DP-SGD (high utility) & 2.2\% & 9.5\% & & 91.13\% ($\pm 0.03$) & 91.12\% ($\pm 0.02$) \\
    DP-SGD (very high utility) & 4.8\% & 63.2\% & & 91.89\% ($\pm 0.03$) & 91.79\% ($\pm 0.03$) \\
    Undefended & 13.4\% & 100.0\% & & 94.52\% ($\pm 0.02$) & 94.10\% ($\pm 0.02$) \\
    \bottomrule
    \end{tabular}
\end{table*}

\section{Experimental Details}
\subsection{Experimental Details in \texorpdfstring{\cref{fig:teaser}}{Figure~\ref{fig:teaser}}}
\label{app:details_for_teaser}

\Cref{fig:teaser} compares typical evaluations of membership privacy defenses
to our proposed protocol.
The bars labeled ``Original'' indicate the TPR@0.1\% FPR of LiRA on a population-level,
while ``Ours'' corresponds to a sample-level evaluation using adaptive attacks (see \cref{ssec:experiments_attacks})
and strong canaries (see \cref{tab:canaries}).
We otherwise use the same experimental setup as in the rest of this paper;
see \cref{ssec:experiments_setup} for details.

For brevity, we only display the high utility DP-SGD baseline,
and skip the medium utility baseline (which is more private).
Similarly, for SSL, we only show the results from a white-box attack on SimCLR,
and omit the (stronger) results for MoCo (see~\cref{fig:ssl_ood_moco}).

We list the full TPR@0.1\% FPR values and test accuracy
in \cref{tab:metrics} for completeness.
Note that our canary choices and our audit setup only marginally affect test accuracy,
but uncover significantly higher privacy leakage.

\subsection{Experimental Details in \texorpdfstring{\cref{fig:average_vs_sample}}{Figure~\ref{fig:average_vs_sample}}}
\label{app:details_for_per_sample_tpr}

The experiments in \cref{fig:average_vs_sample} use 20{,}000 shadow models
trained on \cifar{} without any defense.
Each model randomly includes or excludes every \cifar{} sample in its training data
such that each sample is a member in exactly half (10{,}000) of the models.

To obtain comparable results for population-level and sample-level evaluations,
we first randomly select 64 models as shadow models,
and use the remaining 19{,}936 as victim models.

For every \cifar{} sample $x$,
we then use the 64 shadow models as in the standard LiRA attack to calculate
member and non-member score distributions
$\mathcal{N}(\mu_{x,\textrm{in}}, \sigma_{x,\textrm{in}}^2)$
and $\mathcal{N}(\mu_{x,\textrm{out}}, \sigma_{x,\textrm{out}}^2)$, respectively,
and obtain test scores $\mathcal{A}(f, x)$ for every victim model $f$.
The only difference between the population-level and sample-level metrics is how we aggregate
those test scores.

\myparagraph*{Population-level}
For the population-level evaluation in \cref{fig:per_sample_tpr},
we calculate an ROC curve over test scores $\mathcal{A}(f, \cdot)$ \emph{for each victim model $f$ individually}.
This hence results in 19{,}936 population-level ROC curves,
each over 50{,}000 test scores per victim $f$.
We then determine the TPR@0.1\% FPR for each individual per-model ROC curve,
and report the average over all victim models.
This corresponds to 19{,}936 population-level evaluations as done in previous work;
we report the mean to control for randomness in different victim models.

\myparagraph*{Sample-level}
For sample-level evaluations,
we instead calculate a ROC curve \emph{for each sample $x$ individually},
aggregating the test scores $\mathcal{A}(\cdot, x)$ from all 19{,}936 victim models.
This hence results in 50{,}000 sample-level ROC curves,
each based on 19{,}936 test scores.
We report the TPR@0.1\% FPR for each sample's curve in \cref{fig:per_sample_tpr},
and the full curve of the most vulnerable sample in \cref{fig:roc_common_max_ours}.

\myparagraph*{Top-500 most vulnerable samples and 500 canaries}
First, for the top-500 most vulnerable samples in \cref{fig:roc_common_max_ours},
we determine the 500 samples with the highest sample-level TPR@0.1\%,
and aggregate their test scores on all victim models
(resulting in an ROC curve over $500 \cdot 19{,}936$ test scores).
Second, we use mislabeled samples as the 500 canaries in \cref{fig:roc_common_max_ours}.
We audit those canaries using the same setup as for all case studies in this work
(see \cref{ssec:experiments_setup}),
but also randomly vary the membership of non-audit samples between shadow models
(i.e., include or exclude \emph{every} \cifar{} sample in each shadow model
such that every sample is in exactly half of the models' training data)
to yield a comparable setting.

\subsection{Validation of Our Auditing Setup}
\label{app:aux_validate_loo}
The goal of our evaluation protocol is to mimic realistic model deployments.
However, most existing evaluations vary the membership of all samples in a dataset.
For \cifar{}, this yields 25k training samples in expectation---underestimating
utility, and likely increasing memorization.
We hence use an approach similar to \citet{steinke2023auditing}:
audit only a small subset of the training data,
and always include all other samples in the training data.

While our approach results in realistic models,
it yields a stronger adversary that knows almost all training data.
In the extreme case of a single audit sample,
such an adversary might even reconstruct that sample's features~\cite{ye2023loodistinguishability}.
We thus verify that our approach yields ROC curves comparable to previous methodology.

As in our case studies, we train 64 models,
and attack a small ``audit'' subset of \cifar{} using LiRA in a leave-one-out fashion.
We compare the effects of varying and fixing the membership of non-audit samples as follows:
\begin{enumerate}
    \item Varying membership:
    Resample the training set membership of all \cifar{} samples for each shadow model.
    \item Fixed membership:
    Resample only the membership of audit samples between shadow models,
    and use the same fixed (random) membership for non-audit samples.
\end{enumerate}
Note that both approaches yield an expected training set size of 25k samples;
the only difference is whether non-audit samples are the same for different shadow models.
Varying membership corresponds to most existing evaluations,
while fixed membership mimics our procedure.
For a full picture, we consider both 500 audit samples as in our case studies,
and audit sets of size 250,
proportional to half of \cifar.

\begin{figure}[b]
    \centering
    \includegraphics[width=\figonecolwidth]{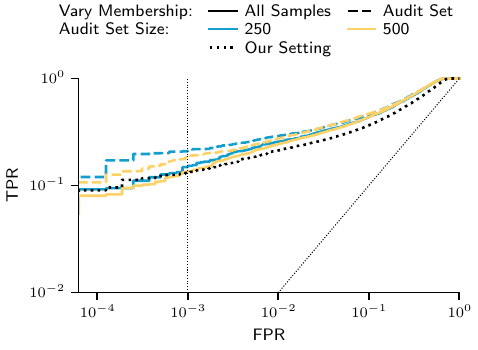}
    \caption{
    \captitle{Our threat model yields comparable privacy estimates to prior work.}
    Recall that prior work trains shadow models by randomly sampling the membership of each dataset point.
    In contrast, we only sample the images that we audit, and keep the remaining parts of the training set fixed.
    While this gives the attacker more power than in prior evaluations,
    we show this has a negligible effect on evaluation outcomes.
    For audit sets of size either 250 or 500,
    we compare the setting where we resample each image (solid lines),
    or where we only resample the audit set
    and keep $\approx 25{,}000$ images fixed (dashed lines),
    and our audit setting (dotted line) where we just sample the audit set,
    and always include the remaining \cifar{} training data (49,500 images).
    Our approach never overestimates privacy leakage.
    }
    \label{fig:validate_loo}
\end{figure}

The results in \cref{fig:validate_loo} show that using the same non-audit samples
in all shadow models (dashed lines) has mild effects.
While the TPR@0.1\% FPR minimally increases
compared to varying the membership of all 50k samples in every shadow model (solid lines),
the difference is negligible compared to the effects of different attacks or canaries.
Considering the setting in our case studies (dotted line),
i.e., including all 49.5k non-audit \cifar{} samples in the training data of all shadow models,
we find that the corresponding ROC curve matches or lies below all four other settings.
Hence, our evaluation protocol allows us to judge a defense's real-world privacy-utility tradeoff
without inflating privacy leakage.

\subsection{Defense-Specific Hyperparameters and Implementation Details}
\label{app:hps}

\begin{figure*}[t]
    \begin{subfigure}[t]{0.48\linewidth}
        \centering
        \includegraphics[width=\figonecolwidth]{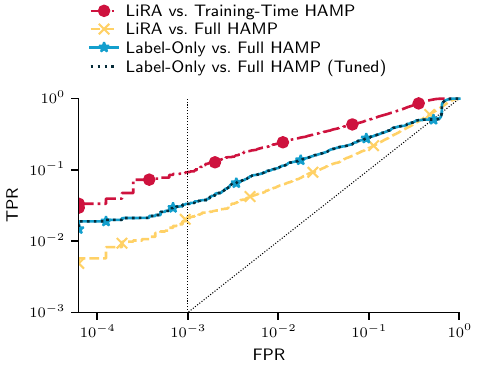}
        \caption{
        Full HAMP ROC curves on a \emph{population-level} (original audit set).
        }
        \label{fig:hamp_full_clean}
    \end{subfigure}
    \hfill
    \begin{subfigure}[t]{0.48\linewidth}
        \centering
        \includegraphics[width=\figonecolwidth]{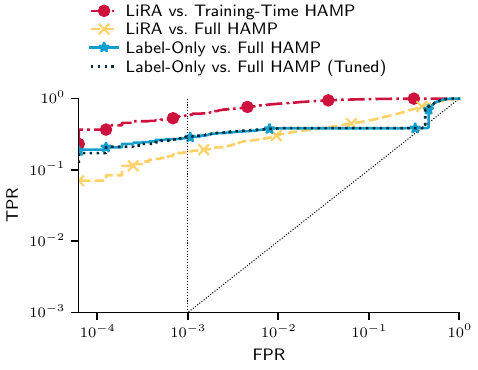}
        \caption{
        Full HAMP ROC curves on \emph{mislabeled canaries}.
        }
        \label{fig:hamp_full_label_noise}
    \end{subfigure}
    \vspace{-0.75em}
    \caption{
        \captitle{Full results for HAMP (a) across the dataset and (b) on mislabeled canaries.}
        In both instances,
        HAMP's training-time defense alone (red) does provide moderate privacy,
        but renders models highly private against LiRA
        when combined with the test-time defense (yellow).
        Yet, our simple label-only attack (blue) successfully
        circumvents the test-time defense.
        We further tune the hyperparameters of the label-only attack (black);
        this yields the default parameters for the population-level evaluation,
        and only marginally influences the ROC curve for the sample-level evaluation on mislabeled canaries.
    }
\end{figure*}

\myparagraph*{HAMP}
We tune the two hyperparameters that directly control the privacy-utility tradeoff
(entropy threshold and regularization strength),
and otherwise use the same hyperparameters as the original paper~\cite{chen2024hamp},
including the same optimizer, learning rate schedule, number of training epochs,
and not using data augmentation.
As the set of potential privacy hyperparameters in~\cite{chen2024hamp}
yields sub-par privacy in our setting, we consider a logarithmic grid of stronger values:
entropy thresholds in $[0.684, 0.9999]$ and regularization strengths in $[0.001, 5.0]$.
From the Pareto-optimal set,
we fix the largest regularization strength that yields stable results,
and pick the largest entropy threshold, subject to 88\% test accuracy.
This results in a regularization strength of $0.005$
and an entropy threshold of $0.9996837722339832$.
For our simple label-only attack, we do \emph{not} tune the ridge regularization strength
of the logistic regression classifiers
(because tuning would require us to train a separate set of shadow models),
and use a default value of $1.0$ instead.
However, we find that even tuning the ridge regularization strength
directly on the victim models does not significantly affect the TPR@0.1\% FPR;
see \cref{app:aux_hamp}.

\myparagraph*{RelaxLoss}
As for HAMP, we tune the loss threshold (as it directly controls the privacy-utility tradeoff),
but otherwise use the same hyperparameters as the original paper~\cite{chen2022relaxloss}.
In particular, we also omit data augmentation,
and restrict posterior flattening only to misclassified samples;
we resolve ambiguities in the original paper by following the authors' implementation.\footnote{
\url{https://github.com/DingfanChen/RelaxLoss/}
}
To find the optimal loss threshold in our setting,
we search a logarithmic grid of values in $[0.0100, 2.1813]$,
and pick the largest threshold that yields at least 88\% \cifar{} test accuracy;
this yields a loss threshold of $0.5946$.

\myparagraph*{SELENA}
Since it is unclear how the number of queries and ensemble members affect privacy,
we use the values proposed by SELENA's authors, that is, performing 10 queries over 25 models.
We further use the same training procedure and hyperparameters as the original paper~\cite{tang2022selena}.

\begin{figure*}[t]
    \begin{subfigure}[t]{\figreducedwidth}
        \centering
        \includegraphics[width=\figreducedwidth]{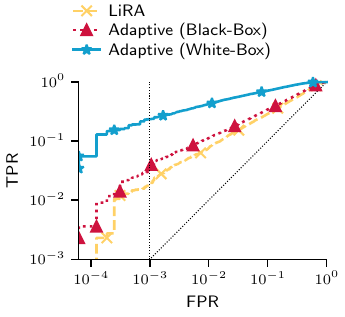}
        \caption{\subcaptitle{Our adaptive attack on the contrastive loss increases privacy leakage for models trained with MoCo, especially in the white-box setting.}}
        \label{fig:ssl_attack_moco}
    \end{subfigure}
    \hfill%
    \begin{subfigure}[b]{0.5\textwidth+0.5\figreducedwidth}
        \begin{subfigure}[t]{\textwidth}
            \centering
            \includegraphics[width=2\figreducedwidth]{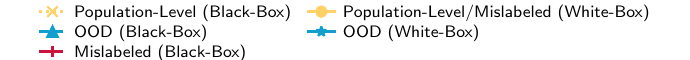}
        \end{subfigure}
        \begin{subfigure}[b]{\textwidth}
            \begin{subfigure}[t]{\figreducedwidth}
                \centering
                \includegraphics[width=\figreducedwidth]{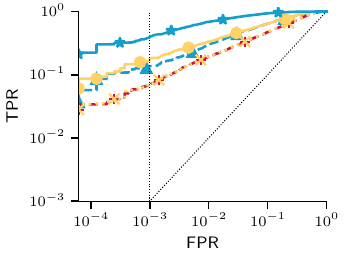}
                \caption{
                Full \emph{SimCLR} ROC curves on OOD canaries.
                }
                \label{fig:ssl_ood_simclr}
            \end{subfigure}
            \hfill%
            \begin{subfigure}[t]{\figreducedwidth}
                \centering
                \includegraphics[width=\figreducedwidth]{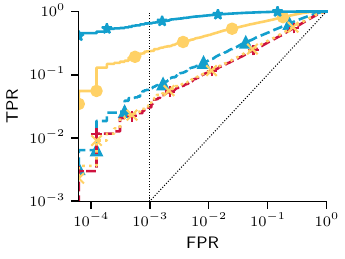}
                \caption{
                Full \emph{MoCo} ROC curves on OOD canaries.
                }
                \label{fig:ssl_ood_moco}
            \end{subfigure}
        \end{subfigure}
    \end{subfigure}
    \vspace{-0.75em}
    \caption{
        \captitle{Additional SSL results.}
        We show the attack success of our adaptive attack for MoCo on a population-level in (a),
        and the full ROC curves on OOD data for (b) SimCLR and (c) MoCO.
    }
\end{figure*}

\myparagraph*{SSL}
For both SimCLR and MoCo, we train an encoder with feature dimension 128 for 800 epochs,
and then fit a linear classifier for an additional 100 epochs while fixing the encoder.
Encoder training, uses a batch size of 512,
momentum 0.9, weight decay 0.0005, and a learning rate of 0.06.
For the linera classifier,
we use a cross-entropy loss, a learning rate of 0.5 and batch size 256.
MoCo additionally relies on a queue during training, we set the size to 4096.

\myparagraph*{DFKD}
Following the setting in~\cite{fang2022up},
we find that only using the ``BN'' loss
(i.e., matching the batch-normalization statistics of the teacher model, widely used in many DFKD methods)
yields a sufficiently high accuracy of at least 88\%.
Therefore, for the sake of simplicity, we only employ that loss for the generation of synthetic data.
This approach facilitates the generalization of our evaluation to numerous other DFKD methods.
Apart from that, we perform for 240 iterations:
In each iteration, we first generate 256 new images, obtain teacher model predictions,
and store the result into a ``memory bank''.
We then train the student model for 5 epochs on the full memory bank.

\myparagraph*{Undefended}
For the undefended baseline in \cref{fig:dpsgd_comparison},
we aim to mimic the hyperparameters of defenses in our case studies.
Concretely, we train WRN16-4 models using SGD with momentum $0.9$ and weight decay $0.0005$,
batch size $256$,
and typical data augmentation (random horizontal flips and shifts of up to 4 pixels).
We optimize for $200$ epochs with a base learning rate of $0.1$;
we linearly warm up the learning rate during the first epoch,
and then decay the learning rate with a factor of $0.2$ at epochs $60$, $120$, and $160$.

\subsection{Canaries}

Table~\ref{tab:canaries} summarizes the adaptive canaries that we use to audit each defense in our study.

\canaryChoicesTable

\section{Additional Experiments}

\subsection{Investigating Adaptive Attacks and OOD Effectiveness for SSL}
\label{app:aux_ssl}

\begin{figure}[t]
    \centering
    \includegraphics[width=\columnwidth]{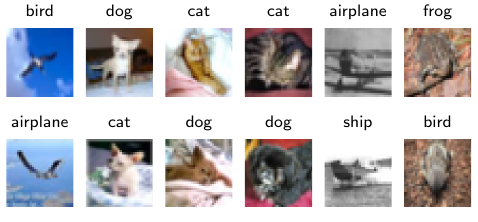}
    \caption{\captitle{Examples of near-duplicates with different labels in \cifar{}.}
    Our canary set for the SELENA defense mimics such samples.
    We highlight a selection of samples (top) and their nearest neighbors (bottom),
    where the two samples have different labels.
    }
    \label{fig:different_labels_nns}
\end{figure}
\begin{figure*}[t]
    \begin{subfigure}[t]{0.48\linewidth}
        \centering
        \includegraphics[width=\figonecolwidth]{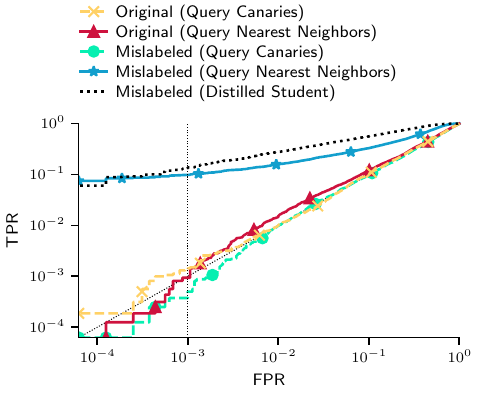}
        \caption{
        \subcaptitle{Naturally occurring near-duplicates in \cifar{} render mislabeled samples
        more vulnerable for SELENA.}
        We directly attack SELENA's ensemble mechanism (\sai{})
        on either original or mislabeled audit samples.
        If we directly query the audit samples,
        the attack never performs better than random guessing.
        However, if we attack a mislabeled sample $(x, \hat{y})$
        by querying its nearest neighbor $x'$
        (the most similar image in the non-audit part of \cifar{}),
        the resulting ROC curve is close to the one of the final distilled student.
        }
        \label{fig:selena_sai_nn}
    \end{subfigure}
    \hfill
    \begin{subfigure}[t]{0.48\linewidth}
        \centering
        \includegraphics[width=\figonecolwidth]{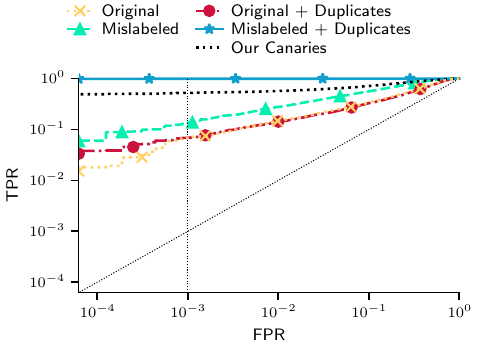}
        \caption{
        \subcaptitle{Including copies of mislabeled canaries in the training data completely breaks SELENA's privacy.}
        We compare SELENA evaluations using original and mislabeled audit samples,
        once as-is, and once by including a copy of the audit set in every model's training data.
        The latter enables close-to-perfect membership inference,
        almost doubling the TPR@0.1\% FPR compared to our proposed canaries---albeit in
        the stronger ``privacy poisoning'' threat model
        introduced in \citet{tramer2022truthserum}.
        }
        \label{fig:selena_student_full}
    \end{subfigure}
    \vspace{-0.75em}
    \caption{
        \captitle{Additional results for privacy leakage in SELENA.}
        In (a), we directly attack SELENA's first defense stage (\sai{}),
        querying either target samples directly, or their nearest neighbor
        in the non-audit data.
        In (b), we consider a slightly stronger threat model,
        where mislabeled duplicates enable
        almost perfect membership inference---even on the full defense.
    }
\end{figure*}

In this section, we present the full results and additional details for both SimCLR and MoCo
in white-box and black-box settings.

\myparagraph*{Adaptive attacks on a population level}
\Cref{fig:ssl_attack_moco} shows the performance of our adaptive attack for MoCo on a population level
(analogous to the SimCLR results in \cref{fig:ssl_attack_simclr}).
We again find that using our adaptive score $\mistat(f; x)$ in LiRA
mildly increases the TPR@0.1\% FPR over standard confidence-based scores in a black-box setting
(2.0\% to 3.6\%), and significantly in a white-box setting (to 23.6\%, more than an $11\times$ increase).

\myparagraph*{Adaptive attacks on OOD canaries}
\cref{fig:ssl_ood_simclr,fig:ssl_ood_moco} depict the full ROC curves
of our adaptive attacks on both SSL defenses,
comparing different types of canaries.
Given that labels neither influence the SSL encoders
nor our white-box attack,
the ROC curves for mislabeled samples and the original audit set
are identical in the white-box setting.
Furthermore, in a black-box setting,
mislabeled samples even yield slightly reduce TPR values.
In contrast, we find that OOD data is a strong canary choice,
since such outliers are significantly more vulnerable for both SSL methods
and threat models.

\subsection{Full HAMP Results}
\label{app:aux_hamp}

As discussed in \cref{ssec:experiments_attacks},
HAMP's test-time defense provides strong privacy
against confidence-based attacks such as LiRA.
The full population-level results in \cref{fig:hamp_full_clean} confirm that,
while the training-time defense alone moderate protects privacy,
the test-time defense reduces privacy leakage by an order of magnitude.
Notably, LiRA achieves a TPR@0.1\% FPR of only 2.1\%---worse than against our high utility DP-SGD baseline!
Yet, our simple label-only attack undoes part of that protection.

For mislabeled canaries, the differences are even more pronounced:
as seen in \cref{fig:hamp_full_label_noise},
our label-only attack increases the TPR@0.1\% FPR by over ten percentage points.
Nevertheless, there is still over a $2\times$ difference between our label-only attack
and LiRA directly targeting the training-time defense.
We suspect that stronger (and more expensive) label-only attacks
can close this gap.

Finally, note that in both cases,
tuning the hyperparameters of our label-only attack only marginally
influences the TPR@0.1\% FPR compared to using default values;
for population-level evaluations, the default and tuned hyperparameters even coincide.

\subsection{Disentangling Privacy Leakage of SELENA}
\label{app:aux_selena}

\myparagraph*{Mislabeled near-duplicates in \cifar{}}
We base our canary choice on the intuition that certain mislabeled/ambiguous \cifar{}
samples leak privacy through a near-duplicate in the training data.
To show that such samples indeed exist,
we calculate OpenCLIP embeddings\footnote{
\url{https://github.com/mlfoundations/open_clip/},
model \texttt{ViT-SO400M-14-SigLIP-384}
pretrained on the \texttt{webli} dataset.
}
of all \cifar{} samples,
and use the pairwise cosine similarity
to determine each sample's nearest neighbor with a different label.
We then inspect the pairs with the highest cosine similarity
and plot a selection in \cref{fig:different_labels_nns}.

This process reveals multiple samples that match our hypothesis,
many with correct labels, and some mislabeled.
For example, we identify an image of a bird that closely
resembles a different bird labeled ``airplane'',
or an image of a Sphynx cat that resembles a different image of a small dog.
Since we use CLIP embeddings to identify those examples,
we argue they are similar not only visually,
but also in terms of neural network features.
Further note that our goal is to identify the \emph{most vulnerable sample} in \cifar{};
hence, while the selection in \cref{fig:different_labels_nns} is small,
a single example suffices.

\myparagraph*{Investigating SELENA's ensemble mechanism}
SELENA's first stage, called ``\sai{}'', is the defense's main privacy mechanism.
Given an ensemble of models $f_i$ and a query sample $x$,
\sai{} aggregates predictions only from models not trained on $x$.
Hence, in isolation, predictions on a training member and non-member should be similar.
The second stage, distillation, serves to reduce computational cost during inference,
and to avoid privacy leakage from certain \sai{} edge-cases.
Yet, SELENA seems to leak more privacy of mislabeled samples
than of the same data with original labels---even without explicit duplicates
(see \cref{fig:selena_student_full}).
We hence analyze \sai{} more thoroughly
to better understand SELENA's behavior on those samples.

Concretely, we consider two SELENA ensembles,
one trained with the original 500 audit samples,
and one with 500 mislabeled audit samples (but without adding duplicates).
We then directly attack \sai{} in two ways:
To obtain predictions on a target sample $x$,
we either query \sai{} on $x$ itself,
or on its nearest neighbor (in the non-audit part of \cifar{}).
As before, we use the cosine similarity of OpenCLIP embeddings as a similarity metric,
and calculate a maximum-weight matching between audit and non-audit samples
to ensure unique nearest neighbors.

The results in \cref{fig:selena_sai_nn} provide further evidence that near-duplicates
are responsible for SELENA's privacy leakage on mislabeled samples.
If an attacker directly queries \sai{} on the audit samples,
the resulting ROC curve is close to a random guessing baseline---even for mislabeled samples.
However, if LiRA queries each audit sample's nearest neighbor instead,
the attack achieves a significantly higher TPR.
Notably, the distillation stage of SELENA queries \sai{} on the full training set,
including nearest neighbors of audit samples.
Hence, the privacy leakage persists in the final student model,
thus explaining the matching ROC curves on mislabeled canaries
for attacks on \sai{} and the distilled student.

\myparagraph*{Stronger threat models}
Attacking \sai{} on our canaries yields almost perfect membership inference
(a TPR of 99.7\% at 0.1\% FPR, and 96.3\% at 0\% FPR),
yet, attacking the distilled student on the same canaries
reduces attack success by about half (52.7\% TPR at 0.1\% FPR).
As argued in the main matter, we suspect that the cause is varying membership of near-duplicates.
We hence consider a stronger threat model,
where only the membership of canaries varies,
and near-duplicates are in the training data of all models.

More concretely, we now mislabel \emph{all} 500 original audit samples,
create a copy of the full audit set (including the wrong labels),
and append this copy to the training data of all models.
As a baseline, we also consider the same procedure with a clean audit set;
that is, we use the same 500 audit samples and copies, but all with the correct labels.
The results in \cref{fig:selena_student_full} show that this stronger model is highly effective:
if duplicates of mislabeled canaries are in the training data of all models,
LiRA achieves an almost perfect TPR of 99.7\% at 0.1\% FPR
(and 99\% with zero false positives)---without explicitly exploiting the presence of duplicates.

While those results are impressive,
the threat model is not entirely realistic.
For one, we could not find any pairs of near-duplicate \cifar{} samples
that are both mislabeled (neither with the same or different classes).
What is more, the threat model resembles data poisoning;
for example, the ``Truth Serum'' attack of \citet{tramer2022truthserum}
renders target samples more vulnerable to membership inference
by inserting copies into the training data of a victim model.

\subsection{Pushing DP-SGD Utility}
\label{app:aux_dpsgd}

Given the strong empirical privacy of our medium and high utility DP-SGD baselines
in \cref{ssec:experiments_dpsgd},
we ask if we can push DP-SGD's accuracy even further,
yet maintain reasonable privacy.
We thus continue tuning the high utility baseline
with the goal of reaching around 92\% \cifar{} test accuracy.

However, as \cref{tb:dpsgd_params_full} shows,
we are unable to achieve our goal.
In particular, our best result raises the test accuracy by less than one percentage point
(to 91.79\%), yet exhibits a sample-level TPR@0.1\% FPR of 63.2\%.
Notably, this is the first instance that one of our case studies (SELENA)
yields stronger privacy at a higher utility than DP-SGD
(even though, at over 50\% TPR@0.1\% FPR,
both defenses are unsuitable for critical applications).
Ultimately, it is likely that achieving very high utility while maintaining strong privacy
is unachievable in practice.
although a formal statement remains an open research question.

\begin{table}[t]
    \caption{Full details for DP-SGD baselines on \cifar{}.}
    \label{tb:dpsgd_params_full}
    \begin{tabular}{@{}lrrr@{}}
    \toprule
     & Medium & High & Very High \\ \midrule
    Test accuracy & $88.29\%$ ($\pm 0.03$) & $91.13\%$ ($\pm 0.03$) & $91.79\%$ ($\pm 0.03$) \\
    DP $\dpeps$ ($\dpdelta = 10^{-5}$) & $\approx 3558$ & $\approx 1.8 \cdot 10^8$ & $\approx 1.1 \cdot 10^9$ \\
    Noise multiplier & $0.2$ & $0.00625$ & $0.003125$ \\
    Clipping norm & $1$ & $1$ & $1$ \\
    Batch size & $2048$ & $64$ & $64$ \\
    Training epochs & $200$ & $16$ & $25$ \\
    Learning rate & $4$ & $4$ & $4$ \\
    Augmult & $8$ & $8$ & $8$ \\ \bottomrule
    \end{tabular}
\end{table}

\subsection{Extended Version of \texorpdfstring{\cref{fig:most_vulnerable_examples}}{Figure~\ref{fig:most_vulnerable_examples}}}
\label{app:aux_most_vulnerable}

In \cref{fig:most_vulnerable_examples}, we show a subset of the most vulnerable \cifar{} samples
(for standard training) to highlight different types of potential canaries.
\Cref{fig:most_vulnerable_examples_aux} contains the full list of the 40 most vulnerable samples
and their corresponding sample-level TPR@0.1\% FPR.

\begin{figure*}[t]
    \centering
    \includegraphics[width=\figtwocolwidth]{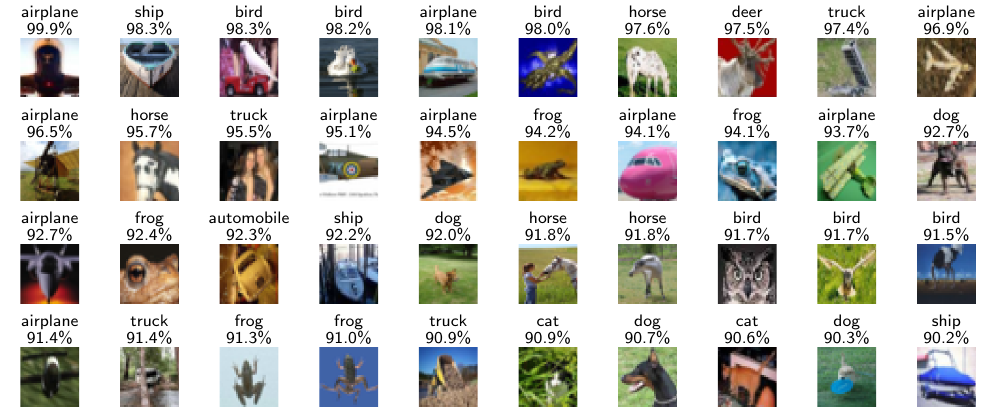}
    \caption{\captitle{The 40 \cifar{} samples most vulnerable to membership inference,
    in the setting of \cref{fig:per_sample_tpr}.}
    The samples are in order of decreasing privacy leakage.
    Titles above each image indicate the original label in \cifar{},
    and the sample-level TPR@0.1\% FPR.
    }
    \label{fig:most_vulnerable_examples_aux}
\end{figure*}

\fi

\end{document}